# Electron nuclear interactions and electronic structure of spin 3/2 color centers in silicon carbide: A high-field pulse EPR and ENDOR study

Victor A. Soltamov[1,2], Boris V. Yavkin[3], Andrei N. Anisimov[1], Ilia D. Breev[1], Anna P. Bundakova[1], Sergei B. Orlinskii[3], Pavel G. Baranov[1]

[1]Ioffe Institute, Politekhnicheskaya 26, 194021, St. Petersburg, Russia

[2]Experimental Physics VI, Julius-Maximilian University of Würzburg, 97074 Würzburg, Germany

[3]Federal Center of Shared Facilities of Kazan State University, 420008 Kazan, Russia

**Abstract**

High-frequency pulse electron paramagnetic resonance (EPR) and electron nuclear double resonance (ENDOR) were used to clarify the electronic structure of the color centers with an optically induced high-temperature spin-3/2 alignment in hexagonal 4H-, 6H- and rhombic 15R- silicon carbide (SiC) polytypes. The identification is based on resolved ligand hyperfine interactions with carbon and silicon nearest, next nearest and the more distant neighbors and on the determination of the spin state. The ground state and the excited state were demonstrated to have spin S = 3/2. The microscopic model suggested from the EPR and ENDOR results is as follows: a paramagnetic negatively charged silicon vacancy that is noncovalently bonded to a non-paramagnetic neutral carbon vacancy, located on the adjacent site along the SiC symmetry c-axis, i.e., the $V_{Si}^{-}$ - $V_C^0$ model with S = 3/2. A number of spin color centers differing in the parameters of the fine structure in the ground and excited states have been discovered and investigated. At the same time, on the basis of the EPR and ENDOR investigations, signs of the fine structure splitting for all the centers were demonstrated, which made it possible to establish the character of optically induced spin alignment, including the inverse populations of the spin levels for these centers. For comparison, ENDOR studies were performed on a negatively charged silicon vacancy $V_{Si}^{-}$ in a regular defect-free environment and a fundamental difference in the properties for these vacancies $V_{Si}^{-}$ and axial centers $V_{Si}^{-}$ - $V_C^0$ was shown. By controlling the neutron irradiation fluence, the spin color centers concentration can be varied over several orders of magnitude down to a single defect level.

## I. INTRODUCTION

Silicon Carbide (SiC) is an advantageous wide-band-gap compound semiconductor for applications in high-frequency, high-temperature, high-power and radiation-resistant electronic devices. The primary defects that can be produced in binary compound SiC (they are often called color centers), are vacancies, interstitials and antisites. In contrast to silicon [1] the primary defects in SiC are stable at, and even far above, room temperature. These primary defects are present at the various sites in the different polytypes that arise from differences in the stacking sequence of the Si and C layers. The cubic 3C-SiC, hexagonal 4H-, 6H-SiC and rhombic 15R-SiC polytypes are the most common and most appropriate for applications. In 3C-SiC only one cubic sublattice site is present. In 4H-SiC two non-equivalent crystallographic positions exist, one hexagonal and one quasi-cubic site, called $h$ and $k$, respectively. In 6H-SiC three non-equivalent positions are formed, one hexagonal and two quasi-cubic ones, called $h$, $k_1$ and $k_2$. In 15R-SiC five non-equivalent positions are formed, two hexagonal and three quasi-cubic ones, called $h_1$, $h_2$, $k_1$, $k_2$ and $k_3$. The nearest neighbors (the first shell) are nearly tetrahedrally oriented for all sites, but the second shells are different for the hexagonal and quasi-cubic sites. The quasi-cubic sites differ in the third shell of neighbors.

Atomic-scale color centers in bulk and nanocrystalline SiC are promising for quantum information processing, photonics and sensing at ambient conditions. Their spin state can be initialized, manipulated and readout by means of optically detected magnetic resonance (ODMR). Until recently, practical applications of semiconductors have been associated with using of defect ensembles. The unique quantum properties of the nitrogen–vacancy (NV) color center in diamond [2] have motivated efforts to find defects with similar properties in SiC, which can extend the functionality of spin color centers not available to the diamond. Such systems are the most prominent objects for applications in new generation of supersensitive magnetometers, biosensors, single photon sources [2, 3, 4, 5-8]. The diamond NV defect is in many ways the ideal qubit, but it is currently quite difficult to fabricate devices from diamond. It remains difficult to gate these defects electrically. A search to find defects with even more potential (“better than excellent”) has now been launched [9-14]. Unique chemical, electrical, optical and mechanical properties make this material very attractive for applications under extreme conditions and can open up a whole new world of scientific applications in spintronics.

A convincing point with SiC is that the stable spinless nuclear isotopes guarantee long dephasing times. Unusual polarization properties of various vacancy related centers in SiC (labelled as P3, P5, P6 and P7) were observed by means of electron paramagnetic resonance (EPR) under optical excitation and reported for the first time in the works of Refs. [15, 16], later in Refs. [17-22, 9, 10].

One of the main questions was to establish whether the observed EPR spectra belong to the ground or to excited state. EPR experiments performed at high frequency and at very low temperatures in darkness excluded the possibility of thermal or optically excited states and as a result it was proved that the EPR spectra of P3, P5, P6 and P7 defects belong to the ground state for all the defects [9, 10, 22]. It has been shown that there are at least two families of color centers in SiC, which have the property of optical alignment of the spin levels and allows a spin manipulation at ambient conditions. (i) Family of silicon-carbon divacancy of the neighboring positions with covalent molecular bond and having a triplet ground state (S = 1). The symmetry of these centers is due to the direction of connection between the silicon and the carbon, zero-field splitting for these centers as in the case of NV-center in diamond is in the gigahertz range. (ii) Family of silicon-vacancy related centers having quadruplet ground and excited states (S = 3/2). The recent experiments demonstrated [9, 10, 13, 20, 22-49] that several highly controllable defects exist in SiC, and some of them can be spin manipulated at room temperature or even higher.

Silicon vacancy related color centers in SiC are demonstrated to be a promising quantum system for single-spin and single-photon spectroscopy. It is assumed that spin-3/2 centers are designated as the corresponding zero-phonon lines (ZFLs): V1, V2, V3, V4. Table I presents the characteristics of spin 3/2 color centers in three polytypes of silicon carbide 4H-SiC, 6H-SiC, and 15R-SiC: zero-phonon line energy/wavelength at 10 K; values of zero-field splitting (ZFS) $\Delta$ ($\Delta=2|D|$) and g-factor of the each center at room temperature (RT).

Zero-field ODMR shows the possibility to manipulate of the ground state spin population by applying radiofrequency (RF) field which is compatible with NMR techniques and using the infrared optical pumping which is compatible with optical fibers and band of transparency of living matter. These altogether make spin 3/2 $V_{Si}$-related defects in SiC very favorable candidate for spintronics, quantum information processing, magnetometry and thermometry.

TABLE I. Zero-phonon line energy/wavelength at 10 K; values of zero-field splitting Δ (Δ=2|D|) and g-factor of the each center at room temperature for family of V1, V2, V3, and V4 spin 3/2 color centers in the crystal lattices of the 4H-SiC, 6H-SiC, and 15R-SiC.

| Polytype | 4H-SiC | | 6H-SiC | | | 15R-SiC | | |
|---|---|---|---|---|---|---|---|---|
| Zero-phonon line | V1 | V2 | V1 | V2 | V3 | V2 | V3 | V4 |
| E, eV/λ, nm | 1.438/862 | 1.352/917 | 1.433/865 | 1.397/887 | 1.368/906 | 1.399/886.5 | 1.372/904 | 1.352/917 |
| Δ MHz/ $10^{-4}$ $cm^{-1}$ | 39/13 | 66/22 | 27/9 | 128/42.7 | 27/9 | 139.2/46.4 | 11.6/3.87 | 50.2/16.7 |
| D MHz/ $10^{-4}$ $cm^{-1}$ | 19.5/6.5 | 33/11 | -13.5/-4.5 | 64/21.35 | -13.5/-4/5 | 69.6/23.2 | -5.8/-1.94 | 25.1/8.35 |
| g-factor | 2.0032 | 2.0032 | 2.0032 | 2.0032 | 2.0032 | 2.005(1) | 2.005(3) | 2.005(3) |

Recently, the existence of NV centers in SiC with S=1 has been demonstrated in [50-53] and can be expected to hold similar promising properties as NV centers in diamond and divacancies in SiC. The $N_C$-$V_{Si}$ centers have been identified in different (3C, 4H, 6H) polytypes of SiC. In the negative charge state, they are spin S=1 centers with optical properties shifted to the NIR region (around 1200-nm wavelength).

At the initial stage of the study of radiation defects, EPR spectra of silicon vacancies in the negatively charged state ($V_{Si}^-$) were found in a regular (defect-free) environment [49]. An important circumstance was that the parameters of the spin Hamiltonian of these vacancies depended little on the polytype of silicon carbide (cubic, hexagonal or rhombic) and the vacancy position (cubic or hexagonal) in the crystal lattice. It should be emphasized that the ZFS for these vacancies, which have an electron spin S = 3/2, is practically zero. Careful studies of g factors by the high-frequency EPR method [22], discovered an extremely small difference in g factors for hexagonal and cubic vacancy positions in 4H-SiC which clearly demonstrated the presence of two positions for the $V_{Si}^-$ in a regular, defect-free environment.

The purpose of this work is to determine hyperfine (HF) interactions with surrounding silicon ($^{29}$Si) and carbon ($^{13}$C) atoms for silicon-vacancy related S=3/2 centers, and thereby find the distribution of the spin density at these atoms and argue for a particular model. It is important to emphasize that in a number of algorithms on the application of spin centers in SiC in quantum processing, the possibility of using nuclear spins of $^{29}$Si and $^{13}$C as a long-term memory is considered. For these purposes, naturally, information is needed on hyperfine interactions with these nuclei, including remote nuclei from the site of the localization of the spin center. This information was obtained using the ENDOR research in this paper.

## II. EXPERIMENTAL

Crystal of 4H-SiC, 6H-SiC and 15R-SiC polytypes were grown by the sublimation technique [54] in vacuum at temperatures between 1700 and 1750 °C and with concentrations of uncompensated nitrogen (N) donors in the range $10^{16}$–$10^{17}$ $cm^{-3}$. These samples have been irradiated with fast neutrons to a dose up to $10^{18}$ $cm^{-2}$ or electrons with the energy of 1–2 MeV with doses ranging from $10^{15}$ $cm^{-2}$ to $10^{16}$ $cm^{-2}$. No annealing treatment was applied. Typical concentration of $V_{Si}$-related color centers was ~$10^{15}$ $cm^{-3}$. The epitaxial SiC layers were also investigated. In addition, 6H-SiC samples of high crystalline quality have been grown by the Lely method.

The EPR and ENDOR spectra were detected at X- (9.3 GHz) and W- (95 GHz) bands on a continuous wave (cw) and pulse (electron spin echo - ESE) spectrometers in the temperature range of 1.2–300 K. The samples, in the shape of platelets, had dimension of about 3×4×0.4 $mm^3$ for X-band and sizes of 0.3×0.4×0.4 $mm^3$ for W-band EPR experiments. The orientation study was facilitated by the possibility to mount the crystal with a high precision owing to the fact that the crystal was cut perpendicular to the c axis and to the fact that the cleaved edge of the sample allowed for a precise rotation in the {11-20} plane. A diode laser operating at 780 nm or 808 nm is used to excite all types of spin color centers directly in the spectrometer cavity through phonon-assisted absorption. All calculations were made using the computer package VISUAL EPR written by V. Grachev, which performs numerical diagonalization of the spin Hamiltonian matrix. [55]

Spin color centers in SiC crystals, intended for EPR and ENDOR measurements, were investigated by the optically detected magnetic resonance method on a confocal microscope combined with a magnetic resonance spectrometer with the ability to control the polytypic composition of the crystal by Raman measurements.

## III RESULTS AND DISCUSSION

A distinctive feature of silicon vacancies in negatively charged state in a regular environment defect-free is the absence of associated photoluminescence. The family of vacancy-related color centers with spin S = 3/2 and axial symmetry along the c axis of hexagonal or orthorhombic SiC polytype possesses completely different properties. The parameter D differs from zero, and has for different centers, both a positive and a negative sign. For these centers, extremely efficient photoluminescence is observed in the near-IR region with a quantum yield close to unity and, most importantly, with an effective mechanism for aligning the populations of the spin sublevels under the effect of optical excitation at room temperatures and above.

To construct a system of energy levels of the spin center, it is necessary to determine the sign of the ZFS from EPR experiments. The EPR spectra of the spin centers in SiC can be described by a spin Hamiltonian of the form

$$\hat{H} = \mu_B \vec{B} \cdot \vec{g} \cdot \vec{S} + \vec{S} \cdot \vec{D} \cdot \vec{S} \,. \quad (1)$$

Here, $\vec{S}$ is the electron spin operator with S=3/2 for color centers. The g-tensor $\vec{g}$ reflects trigonal symmetry with the principal values $g_{||}$ and $g_{\perp}$ corresponding to the directions parallel and perpendicular to the hexagonal *c*-axis, $\mu_B$ is the Bohr magneton. The first term describes the electron Zeeman interaction and the second term reflects the fine structure for $S$>1/2 which for an axial symmetry can be written as $D[\hat{S}_z^2 - 1/3S(S+1)]$.

EPR experiments at low temperatures and at a high operating frequency give information about sighn of D parameter. Such experiments were partly carried out by us earlier [22] and new results have been obtained in [57] and the present work. Figure 1 shows the ESE-detected EPR spectra of V2 centers observed at W-band in 4H-SiC (a), 6H-SiC (b) crystals measured at low temperature (1.2 K) and high temperature (190 K and 50 K) at two orientations between the magnetic field B and the crystal c-axis without light excitation (light off) with Boltzmann population distribution. In Fig. 1(c) the EPR spectra of V2 and V1/V3 centers (later on, we will denote these centers in 6H-SiC as V1/V3, since the zero-field splitting is the same, see Table I) observed at X-band in 6H-SiC in orientation B || c at low temperature of 5 K and high temperature of 18 K without light are presented. In addition, the EPR spectra under light excitation are shown at 5 and 60 K. Figure 1(d) shows the ESE-detected EPR spectra of V2 centers observed at W-band in 15R-SiC at three temperatures (50, 10, 8 K) for the orientation close to B || c without light (solid line); dashed line shows for comparison the EPR spectra under 780 nm light excitation at 50 K. One can see that the optical pumping leads to a sharp deviation of the population of the spin sublevels from the Boltzmann distribution, and as a result, the signal intensity increases significantly while the phase of the low-field signal is inverted, Fig. 1 (d) and (c), since instead of absorbing the microwave power, radiation is observed at this transition.

Observation of the strong EPR signals at low temperature and in complete darkness proves that these signals belong to the high-spin ground state. One can see that in the EPR spectra at low temperatures the intensities of the fine-structure components differ strongly due to the extreme difference in the populations of the spin sublevels at this low temperature and the large Zeeman splitting. For the low-temperature spectra at 1.2 K, Fig. 1(a, b), only the high-field line can be observed for $B \,||\, c$ and only the low-field line for $B \perp c$. The same tendency was observed in 15R-SiC, but in these experiments the temperature was not lower 8 K, therefore, the Boltzmann factor was significantly less than in the case of 1.2 K and the low-field EPR line of the V2 centers was also observed, but was significantly less than the high field line. This result allows us to decide that a sign of the fine-structure D parameter of the V2 center in all three polytypes is positive.

The inset in Fig. 1(b) shows the energy level diagram for V2 center with spin S = 3/2 in 6H-SiC and conditionally shows the population of the spin sublevels in accordance with the Boltzmann distribution in the case of low temperatures and large microwave quanta and the corresponding EPR transitions. It can be seen that the intensity of the EPR lines is determined by the level populations, for D> 0, the high-field signal is much more intense than the low-field signal; in the case of high temperatures, the populations of the levels differ insignificantly and, therefore, the high-field and low-field EPR signals are approximately equal. It was possible to compare the intensities of the EPR lines for spin centers with maximal fine structure splittings, labeled V2 in all the crystals and for V1/V3 centers in 6H-SiC. For V1/V3 centers in 6H-SiC, the opposite intensity ratio was observed between the high-field and low-field lines at low temperature, Fig. 1(c), which gives the sign D<0. For the remaining centers in 15R-SiC, as shown below, the signs of the fine structure splitting were determined from the ENDOR data, based on the results of EPR and ENDOR studies of V2 centers in three polytypes of silicon carbide, and V1/V3 centers in 6H-SiC, which can be considered as a starting point for finding the signs of the D parameter for other types of centers: by comparing the EPR and ENDOR results.

In the next stage, the ENDOR spectra of the spin centers V1, V2, V3 in 6H-SiC; V2, V3, V4 in 15R-SiC (the EPR spectrum of V1 centers was not observed in 15R-SiC) and V2 centers in 4H-SiC will be investigated. Figures 2-10 show the results of ENDOR measurements for different axial spin-3/2 centers in 6H-SiC, 15R-SiC and 4H-SiC polytypes. Before analyzing ENDOR spectra of spin centers to elucidate hyperfine (HF) interactions with different shells, we consider information that was obtained directly from the hyperfine structure observed in the EPR spectra [56] and are given in Table II.

The HF structure in EPR and ENDOR spectra of the color centers in SiC can be described by the additional terms in the spin Hamiltonian of Eq. (1) in the form

$$\sum_i (\vec{S} \cdot \vec{A}_i \cdot \vec{I}_i - g_{Ni}\mu_N \vec{B} \cdot \vec{I}_i) \quad (2)$$

Here, $\vec{S}$ is the electron spin operator with S=3/2 for color centers and $\vec{I}_i$ represents the nuclear spin operators for $^{29}$Si ($I_{Si}$=1/2) or $^{13}$C ($I_C$=1/2) nuclei located at different neighbor shells of the Si sites and C sites, $g_{Ni}$, is a g factor of nucleus i ($g_N$ is negative for $^{29}$Si and positive for $^{13}$C), $\mu_N$ is the nuclear magneton. The first term reflects the hyperfine interaction where $\overset{\leftrightarrow}{A}_i$ is the tensor of this interaction, which describes the HF interaction with the ith Si or C atoms, located at different neighbor shells of the spin color center. HF interactions in the first and the second shells of the Si vacancy are partly resolved in the EPR spectra. The second term describes the nuclear Zeeman interaction for $^{29}Si$ and $^{13}C$ nuclei. All calculations of the spin Hamiltonian (1, 2) were made using the computer package, which performs numerical diagonalisation of the spin Hamiltonian matrix [55].

ENDOR transition frequencies determined by the selection rules $\Delta M_S = 0$ and $\Delta m_I = \pm 1$ are given by [58]:

$\nu_{ENDORi} = h^{-1}|M_S[a_i + b_i(3\cos^2\theta - 1)] - g_{Ni}\mu_N B|$ (3)

where $a_i$ and $b_i$ are isotropic and anisotropic parts of the HF interaction with the ith nucleus, θ is the angle between the external magnetic field B and the HF interaction tensor, $g_{Ni}\mu_N B/h$ is the Larmor frequency $f_L$. The HF interaction tensor components can be expressed in terms of the isotropic *a* and anisotropic *b* components as $A_{\|}=a+2b$ and $A_{\perp}=a-b$ with axial symmetry around the *p* function axis. Here, $a = (8\pi/3)g_e\mu_B g_N\mu_N|\Psi_{2s}(0)|^2$ and $b=(2/5)\ g_e\mu_B g_N\mu_N\langle r_{2p}^{-3}\rangle$, where $g_e$ is the electronic g factor, and Ψ is unpaired-electron wave function.

Let us single out the main features observed in the ENDOR spectra common to all spin centers and polytypes of SiC.

1. The most intense ENDOR lines are observed in the Larmor frequency region of $^{29}Si$, with signals visible on both sides of the Larmor frequency.

2. The ENDOR signals registered over the low-field (lf) and high-field (hf) EPR lines have mirror symmetry with respect to the Larmor frequency with phase inverting (a small shift of all the lines is due to a small difference in the Larmor frequency owing to different magnetic fields for the EPR transitions).

3. The position of all the ENDOR lines in one spectrum is not symmetric with respect to the Larmor frequency. Equation (3) predicts that interaction with each ith nucleus induces two sets of ENDOR transitions located at the distances of $1/2A_i$ and $3/2A_i$ from the Larmor frequency $f_L$. Thus, in all spectra it is possible to distinguish pairs of lines located at a distance of $1/2A_i$ and $3/2A_i$ from the Larmor frequency on either side of the Larmor frequency, while the constant $A_i$ is the $^{29}Si$ hyperfine interaction constant for each particular silicon atom “i” varies widely and has a different sign for different silicon atoms.

4. Some of the positions of the ENDOR lines depend on the orientation of the crystal in the magnetic field (angle θ), that is, the lines are anisotropic (see Eq. (3)), and some of the lines are practically independent of the orientation of the crystal in the magnetic field, that is, they are almost isotropic. The EPR and ENDOR spectra describe hyperfine interactions, the largest of these interactions lead to splitting in the EPR spectra or to broadening of the EPR lines. An important task is to compare these interactions and identify those that are observed, both in the EPR spectra and in the ENDOR spectra. Such binding research allows, as a rule, to exclude erroneous interpretation of the ENDOR spectra.

The positive sign of the parameter D for V2 centers in 4H-SiC, 6H-SiC and 15R-SiC (Fig. 1a, b, c, d) and the negative sign for V1/V3 centers in 6H-SiC (Fig. 1b, c) were established from EPR spectra. The ENDOR measurements confirmed the EPR data and made it possible to determine the signs of the D parameter for other spin centers that were not found in the EPR studies. Using this information it will be possible to identify the transitions observed in the EPR spectra and, ultimately, from the position of the signals with the inverted phase in the magnetic field to find optically induced populations of spin sublevels in the ground S=3/2 state.

Attention is drawn to the similarity of ENDOR signals for different centers and polytypes: there are always signals with maximum interactions with $^{29}$Si nuclei belonging to the second shell around the silicon vacancy – the next-nearest neighbors NNN ($Si_{NNN}$) and corresponding to almost isotropic satellites that appear in the EPR spectra. In contrast to EPR spectra there is a set interactions in ENDOR signals with closely related parameters, since twelve NNN silicon atoms in this shell are slightly nonequivalent.

In accordance with the identified EPR transitions for V2 centers, due to the known parameter D, for these interactions the HF interaction constant is positive, that is, the sign of the spin density on the silicon nuclei is negative in this case (in view of the negative nuclear g-factor for silicon, $^{29}$Si). This result is in agreement with the theoretical calculation performed for isolated silicon vacancies in a regular defect-free environment $V_{Si}^-$ [49], where a negative sign of the spin density for silicon in the second coordination sphere relative to the silicon vacancy ($Si_{NNN}$) was predicted.

HF interactions with positive constants are also correspond to a whole set of lines located on the same side relative to the Larmor frequency $^{29}$Si as the lines responsible for interactions with twelve silicon atoms $Si_{NNN}$. These lines are naturally attributed to the interaction with the farther spheres of silicon.

The next stage is a consideration of the ENDOR lines located on the opposite side of the $^{29}$Si Larmor frequency and, therefore, having the opposite sign of the HF interaction with $^{29}$Si nuclei, that is, a negative HF interaction value corresponding to a positive spin density on the silicon core. A distinctive feature of this interaction is its relatively large magnitude and sharp anisotropy, which resembles the anisotropy of hyperfine interaction with four carbon nuclei located in the nearest neighborhood (NN) of a silicon vacancy, $C_{NN}(V_{Si})$, and well-studied by EPR methods (see Fig. 3 (b) for 6H-SiC and Fig. 8 (a) for 15R-SiC). This result gives us grounds to assert that these signals belong to four silicon atoms located in the nearest neighborhood (NN) of the carbon vacancy $V_C$, $Si_{NN}(V_C)$, which corresponds to a model of the axial spin-3/2 center previously proposed by us (see book [59] and references therein).

We compare the ENDOR spectra of different centers in different polytypes in order to determine the sign of D at those centers for which it was not possible to determine these signs from the EPR spectra. It is obvious that the signs of the HF interactions, and, consequently, of the spin densities are almost the same for all centers. Thus, the position of the ENDOR lines relative to the $^{29}$Si Larmor frequency (with a frequency of a greater or less Larmor frequency) bears information about the EPR transition (the sign of $M_S$), which ultimately makes it possible to determine the D sign. The D sign uniquely determines the $M_S$ sign for the EPR transitions observed in smaller and larger magnetic fields. After the system of energy levels has been determined and the phase relationships between EPR signals for optically aligned populations of spin levels (the order of transitions in a magnetic field for radiation and absorption of microwave), it is possible to establish a character of optically induced populations of spin sublevels.

### A. Spin 3/2 color centers in 6H-SiC single crystal

Figure 2 shows ESE detected ENDOR spectra at W-band of the V2 and V1/V3 centers in 6H-SiC, recorded in orientation B||c with a scan in a wide frequency range. The ENDOR magnetic fields correspond to the high-field (hf) transitions indicated in optically induced ESE detected EPR spectra shown at the right. The ENDOR lines with the strongest HF interactions (in absolute value) are marked. There is a mirror reflection of the HF interaction lines with different coordination spheres of silicon relative to the Larmor frequency of $^{29}$Si for V2 and V1/V3 centers, which unambiguously indicates opposite signs of fine structure splitting. The insert shows one EPR line of V2 center in 6H-SiC crystal with a natural isotope silicon content and with a modified isotopic composition of silicon (0.7% of $^{29}$Si isotope). The satellite lines observed in the EPR spectra are due to the HF interactions with twelve Si atoms located in the next-nearest-neighbor (NNN) of a silicon vacancy $V_{Si}^{-}$ and these lines correspond to the maximum interactions for the ENDOR lines designated in Fig. 2 as $A_{SiNNN}$ ($Si_{NNN}$ around $V_{Si}^{-}$). Almost isotropic hyperfine interaction with the $^{29}$Si $A_{SiNNN}$ are visible both in EPR and in ENDOR. In EPR spectra the interaction with different $Si_{NNN}$ atoms of the second shell is not distinguishable, whereas in the ENDOR spectra these lines are resolved and are isotropic within the widths of the ENDOR lines. It follows from formula (3) that the HF interaction constants $A_{SiNNN}$ is positive and corresponds a negative spin density (as the nuclear $g$ factor $g_N$ is negative for $^{29}$Si). This observation agrees with the results of the theoretical calculation for isolated silicon vacancy in regular defect-free environment $V_{Si}^{-}$ [49].

Figure 3 shows W-band angular dependence ESE-detected EPR (a) and corresponding ESE-detected ENDOR (b) spectra of the optically aligned V2 color centers in single 6H-SiC crystal. The expanded-scale ESE-detected ENDOR spectrum from Fig. 3 (b) is presented in Fig. 3 (c). The EPR transitions for the low-field (lf) and high-field (hf) magnetic fields are indicated in optically induced ESE spectra. The dotted lines in the Fig. 3 (b, c) correspond to the signals recorded in the additional 6H-SiC sample under the changed experimental conditions, in which ENDOR lines are visible in B||c, which are weakly manifested in the main sample because of blind spot effect (see explanation below). The angular dependencies of the ENDOR lines are highlighted in gray.

Two types of HF interactions were directly observed in the ESE-detected EPR spectra. The first type of the HF interactions occurs with the $^{13}C$ nucleus located in the nearest neighbor (NN) shell to the $V_{Si}^{-}$ site. They are strongly anisotropic and reflect the tetrahedral symmetry of the nuclear spin locations. The inset in Fig. 3 (a) displays a resolved in the EPR spectrum anisotropic hyperfine structure for the NN carbon atoms with respect to negatively charged silicon vacancy $V_{Si}^{-}$; and almost isotropic hyperfine structure for the NNN Si atoms with respect to the negatively charged silicon vacancy $V_{Si}^{-}$. $^{13}C_1$ denotes the interaction with the carbon atom oriented along the c axis and $^{13}C_{2\text{–}4}$ denotes the interactions with atoms located in the basal plane with the bonds inclined by the angle $\theta = 71°$ relative to the c axis. The HF structure arising from such interactions is presented in Table II [56].

There are two types of main HF interactions with silicon $^{29}Si$ nuclei, which differ in magnitude, sign and anisotropy. They can be considered in the framework of the spin center model proposed earlier as interactions with silicon atoms surrounding silicon or carbon vacancy. In Fig. 3 (b), the signals at $f_L$-1/2$|A_{SiNNN}|$ and $f_L$-3/2$|A_{SiNNN}|$ correspond to HF interactions with the NNN Si atoms with respect to the negatively charged silicon vacancy $V_{Si}^{-}$. All the lines are grouped together because of the almost isotropic hyperfine interaction. The light-induced inverse population of the spin sublevels of V2 centers is shown in the inset. In Fig. 3 (c), the signals at $f_L \pm 1/2|A_{SiNN}|$ and $f_L \pm 3/2|A_{SiNN}|$ indicate ENDOR lines corresponding to the HF interactions with the nearest-neighbors (NN) Si atoms with respect to the neutral carbon vacancy $V_C^0$.

One of the drawback of the pulse Mims-type ENDOR [60] is occurrence of blind spot regions [61, 62]. Blind spots show up when the ENDOR transitions induce a shift of frequency equal to $2\pi(n/\tau)$, where n=0, ±1, ±2, …. To avoid the possibility of mission individual lines, one has to take a multitude of ENDOR recordings at various settings of $\tau$ and following reconstruction of the complete ENDOR spectrum from several scans recorded at different values of $\tau$. The low intensity of some lines in ENDOR spectra is a manifestation of the blind spot effect of Mims-type ENDOR (see Fig. 3 and Fig. 4), and is associated with different experimental conditions for recording

ENDOR spectra. Figure 4 demonstrates a manifestation of the blind spot effect of Mims-type ENDOR into experiments and affected the part of spectrum, preventing ENDOR lines at frequencies 29.6 MHz and 31.8 MHz from being observed in two upper plots. Notation of HF interaction $A_1$ was introduced for the ENDOR lines of the nearest-neighbor $Si_{NN}$ atoms around a carbon vacancy $V_C^0$ located along the *c* axis of the tetrahedron around the carbon vacancy, then $A_{2-4}$ correspond to three $Si_{NN}$ atoms in the basal plane with the bonds inclined by the angle $\theta = 71°$ relative to the c axis.

Figure 5 shows an angular dependence of W-band ESE-detected ENDOR for V2 color centers in single 6H-SiC crystal. Hyperfine interactions with the surrounding silicon nuclei $^{29}Si$ (the left-hand side) and carbon $^{13}C$ (right-hand part) are presented. The angular dependencies of the ENDOR lines are highlighted in gray and by dashed lines. If the spin density $\rho_S = \Sigma(|\Psi_{s\uparrow}|^2 - |\Psi_{s\downarrow}|^2)$ is the same in sign (for example, $\rho_S < 0$ as in twelve $Si_{NNN}$ atoms), then the HF structure constant $A_S \propto g_N(^{29}Si) > 0$ and $A_S(^{13}C)<0$ since $g_N(^{29}Si)=-1.1106 <0$ and $g_N(^{13}C)=1.40482 > 0$. As a result of expression $\nu_{ENDOR} = \hbar^{-1}|M_S A - g_N\mu_N B|$, for the same sign of the spin density, the ENDOR lines of $^{29}Si$ and $^{13}C$ are shifted to the same side with respect to the Larmor frequency. For the upper spectrum in Fig. 5, anisotropic signal for $^{13}C$ (about 0.5 MHz) to the right of Larmor frequency $^{13}C$ corresponds to a positive spin density, as well as an anisotropic signal for $^{29}Si$ (about 2.5 MHz) to the right of Larmor frequency $^{29}Si$. We conclude that this is respectively $C_{NNN}$ and $Si_{NN}$ with respect to the neutral carbon vacancy, that is, $C_{NNN}$ ($V_C^0$) and $Si_{NN}(V_C^0)$. The signals to the left of the Larmor frequency of $^{29}Si$ and the Larmor frequency of $^{13}C$ correspond to negative spin density and refer to $Si_{NNN}(V_{Si}^-)$ and $C_{IV}(V_{Si}^-)$.

The surprising result is the observation of practically identical orientational dependences of ENDOR signals from silicon nuclei and carbon nuclei (indicated by dashed lines in Fig. 5), probably located in identical shells relative to the vacancies of silicon and carbon entering the spin center, which may be one more evidence in favor of proposed two-vacancy model.

Figure 6 shows W-band angular dependence ESE-detected EPR (a) and ESE-detected ENDOR (b) spectra of the optically aligned V1/V3 color centers in single 6H-SiC crystal. The expanded-scale ESE-detected ENDOR spectrum from Fig. 6 (b) is presented in Fig. 6 (c). The EPR transitions for the low-field and high-field magnetic fields are indicated in optically induced ESE spectra.

As in the previous figures, in Fig. 6 (b), the signals at $f_L+1/2|A_{SiNNN}|$ and $f_L+3/2|A_{SiNNN}|$ correspond to HF interactions with the next-nearest-neighbors (NNN) Si atoms with respect to the negatively charged silicon vacancy $V_{Si}^-$. The light-induced inverse population of the spin sublevels of V1/V3 centers (D<0) is shown in the top inset. Bottom inset displays expanded-scale ESE-detected ENDOR signals due to an almost isotropic hyperfine interaction with carbon atoms (with

$^{13}C$ nuclei) apparently located in the third coordination sphere with respect to negatively charged silicon vacancy $V_{Si}^{-}$.

The signals at $f_L \pm 1/2|A_{SiNN}|$ and $f_L \pm 3/2|A_{SiNN}|$ in Fig. 6 (c) indicate ENDOR lines corresponding to the HF interactions with the nearest-neighbors Si atoms with respect to the neutral carbon vacancy $V_C^0$. From the ENDOR spectra for carbon $^{13}C$, we obtain the HF interaction constants of the positive sign, which corresponds to a positive spin density at the carbon nuclei (as the nuclear *g* factor $g_N$ is positive for $^{13}C$).

### B. Spin 3/2 color centers in 15R-SiC single crystal

Figure 7 shows ESE detected ENDOR spectra in a wide range of radio-frequencies at W-band of the V2 centers in 15R-SiC, B~|| c for the lf and hf transitions indicated in optically induced ESE detected EPR spectra shown at the right. Triangles denote the ENDOR lines with the strongest HF interactions (in absolute value) and negative HF constants with twelve NNN Si atoms, $Si_{NNN}(V_{Si}^{-})$.

W-band angular dependence ESE-detected EPR (a) and corresponding expanded-scale ESE-detected ENDOR (b) spectra of the V2 color centers in single 15R-SiC crystal are presented in Fig. 8. As in the case of 6H-SiC two types of HF interactions were directly observed in the ESE-detected EPR spectra in 15R-SiC: (1)The HF interactions occurs with the $^{13}C$ nucleus located in the (NN shell to the $V_{Si}^{-}$ site. They are strongly anisotropic and reflect the tetrahedral symmetry of the nuclear spin locations. As an example the HF lines arising from these interactions in 15R-SiC are shown in Fig. 8 (a): $^{13}C_1$ denotes the interaction with the carbon atom oriented along the c axis and $^{13}C_{2-4}$ denotes the interactions with atoms located in the basal plane with the bonds inclined by the angle $\theta = 71°$ relative to the c axis. The HF structure arising from such interactions is presented Table II, which match closely previously reported values for the $V_{Si}^{-}$ centers in 4H-SiC and 6H-SiC [56]. (2) The HF interactions occur with the $^{29}Si$ nucleus located in the NNN shell to the $V_{Si}^{-}$. These interactions (Table II) are shown in the inset in Fig. 8 (a). Variation in the orientation of the magnetic field did not change the line splitting in EPR; only the strong anisotropy of the linewidth was observed.

The angular dependencies of the ENDOR lines for the $Si_{NN}$ atoms around a carbon vacancy $V_C$, are highlighted in gray. Preliminary results on one type of color center in 15R-SiC crystal were presented by us in Ref. [39]. The strong anisotropy of the ENDOR spectra marked in the Fig. 8 (b) as $f_L$-$1/2|A_{SiNN}|$ and $f_L$-$3/2|A_{SiNN}|$ (4Si atoms around $V_C$) is noteworthy.

Figure 9 shows ESE detected ENDOR spectra at W-band on an enlarged scale of the V2, V3 and V4 color centers in 15R-SiC measured at orientation $\theta \approx 10^0$ for the lf and hf transitions indicated in optically induced ESE spectra shown at the right. The insets show the schemes of energy levels and their optically induced populations for the studied color centers. The ENDOR spectra for V2 centers are given for comparison (one of the spectra shown in Fig. 8 is repeated). The inserts show the energy schemes of spin sublevels for V2 centers, and also conditionally show the populations of the spin sublevels at room temperature as a result of optical alignment. It is seen that the levels with $M_S$ = ±3/2 are predominantly populated, that is, taking into account the positive sign of the fine structure (D>0), an optically induced inverse population of the spin sublevels is observed at room temperature even in a zero magnetic field.

Let us consider in more detail the ENDOR spectra of the V3 and V4 centers and on the basis of the ENDOR research and taking into account that only the sign of the fine structure for V2 centers is known from the EPR studies, we shall construct the scheme of the energy levels and population of these levels under the action of optical pumping at room temperature.

As a basis, we take the similarity of the ENDOR spectra for hyperfine interactions with silicon nuclei $^{29}$Si surrounding the silicon vacancy. These interactions differ little for different spin centers with S = 3/2 and, as will be shown later, practically coincide with the ENDOR spectra obtained for an isolated silicon vacancy with zero splitting of the fine structure (D=0). These interactions are characterized by a positive sign of the hyperfine structure constant (i.e., a negative spin density on silicon nuclei), as was shown on the basis of studies of V2 centers, since the sign of D for these centers is determined from the EPR measurements, hence $M_S$ for EPR transitions in high and low fields are identified, which are included in the Eq. (3) for determining the frequencies of the ENDOR lines.

First, we compare the relative position of the EPR lines in the magnetic field and the position of the ENDOR line with respect to the Larmor frequency for the V2 center on one side (which is known) and the V3 and V4 centers under consideration (V1 the center is not considered, since we did not find the EPR spectrum corresponding to the zero-phonon line V1). At the first stage, to identify the ESR transitions, it does not need information on the inverting of ESR signals as a result of optical pumping. This information in the future, after establishing the order of energy levels (sign D), will make it possible to find populations of spin sublevels under the action of optical pumping. Figure 9 shows that the low-field lines V2 and V4 correspond to ENDOR signals for interaction with silicon nuclei with a positive HF constant located in the frequency range above the Larmor frequency, whereas for V2 and V3, the ENDOR signals correspond to the interaction with silicon nuclei with a positive HF interaction constant, located in the range above and below the Larmor frequency. This gives us reason to state that for V2 and V4 low-field and high-field transitions correspond to the same values of $M_S$, that is, in both cases the sign of D is the same (D> 0). For V3 centers, the opposite situation is observed, that is, the sign D is negative (D <0). These results are reflected in the corresponding inserts, where the energies of spin sublevels in a magnetic field for V4 and V3 centers are given.

The population of the spin sublevels, obtained on the basis of the position in the magnetic field of EPR signals with an inverted phase, that is, the signal of microwave radiation (Fig. 9). Now consider the ENDOR signals with a negative HF interaction constant, that is, corresponding to interactions with silicon atoms located in accordance with our model near the carbon vacancy. A distinctive feature of these HF interactions is a strong anisotropy, which resembles the anisotropy of HF interactions with the nearest neighbor carbon atoms (4 atoms, $C_{NN}$) located in the first coordination sphere around the silicon vacancy. This analogy allows us to assert that four silicon atoms are in the nearest environment of the vacancy of carbon $V_C$, $Si_{NN}(V_C)$, just as four carbon atoms are located near the silicon vacancy, $C_{NN}(V_{Si})$, see Fig. 8 (a). In this case, according to our model, the silicon vacancy is negatively charged ($V_{Si}^-$) and is responsible for the total spin S=3/2, the carbon vacancy is neutral ($V_C^0$), so the HF interaction is due to the core polarization mechanism. It is known that the total spin density due to the core polarization can have a positive and negative sign, depending on the distribution of closed electron shells with respect to unpaired electrons (with total spin S=3/2 for V1, V2, V3 or V4 spin center) [63].

**C. Spin 3/2 color centers in 4H-SiC single crystal**

Figure 10 presents ESE detected ENDOR spectra at W-band of the V2 color centers in 4H-SiC measured at orientation $\theta \approx 0^0$ for the low-field (lf) and high-field (hf) transitions indicated in optically induced ESE spectra shown at the right. For comparison, the $^{29}$Si ENDOR spectra of the V4 centers in polytype 15R-SiC and V1/V3 centers in 6H-SiC and $^{13}$C ENDOR spectra of the V2 center in 6H-SiC are shown. The inset shows the level system for the V2 center in 4H-SiC.

Figure 11 shows a cw (a) and ESE (b) W-band spectra (94.9 GHz) of the silicon vacancy in a regular defect-free environment ($V_{Si}^-$) observed at 300 K (a) and 1.2 K (b) in n-irradiated 4H-SiC (dose of $10^{18}$ cm$^{-2}$) for several orientations of the magnetic field with respect to the c axis including the orientations parallel (θ=0°) and perpendicular (θ=90°) to the c axis. The central line and two HF satellites are shown for the h and k sites. The intensity ratio of the central line to that of the satellites corresponds to the interaction with 12 silicon atoms of the second shell. The results of simulation of the EPR (a) and ESE (b) spectra are presented.

In the EPR spectra of the silicon vacancy in a regular defect-free environment ($V_{Si}^-$) in 4H-SiC an anisotropic splitting of the EPR lines is observed. This splitting is assumed to arise from small differences in the g tensor of the quasicubic (k) and hexagonal (h) sites [22]. The g tensor for the k site g(k) is found to be isotropic with g(k)=2.0032 and the g tensor of the h site is found to be slightly anisotropic with $g_{||}(h) = g(k)+0.00004$ and $g_{\perp}(h)=g(k)+0.00002$. Pulsed EPR measurements at 95 GHz confirm that the spin multiplicity of $V_{Si}^{-}$ is S=3/2 in agreement with the results of ENDOR measurements. The fine-structure D parameter for $V_{Si}^{-}$ is close to zero ($D<0.5x10^{-4}$ $cm^{-1}$). It is important to note that no photoluminescence is associated with silicon vacancy in a regular defect-free environment, in contrast to the anisotropic S=3/2 spin centers considered in this work.

EPR spectra of the silicon vacancy in a regular defect-free environment were studied in all the main polytypes of silicon carbide, including cubic. Based on the study of HF interactions with the twelve Si-atoms in the NNN shell in 4H-SiC, it was found that the spin is 3/2 [49]. Additional information on vacancy centers in a defect-free environment was obtained by the ENDOR method on the example of a 15R-SiC crystal. Figure 12 shows ESE detected EPR (a) and ENDOR (b) spectra at W-band of the negatively charged silicon vacancy in regular defect-free environment ($V_{Si}^{-}$ center) in 15R-SiC. It can be seen that these ENDOR spectra differ significantly from those shown in Figures 2-10 and which belong to anisotropic spin S=3/2 color centers. There are no interactions with silicon atoms, which, as we assume, are located around the carbon vacancy entering the spin center. A serious reason for considering the more complex structure of the axial spin-3/2 center is the absence of such centers in cubic silicon carbide, in which there are no “zigzags” in the structure of the crystal lattice. The presence of “zigzag” is necessary for the formation of the centers $V_{Si}^{-}$ - $V_C^0$ shown in Fig. 13. Note, that isolated silicon vacancies in regular environment are manifested in spin-dependent recombination in experiments on electrical detection of magnetic resonance (EDMR) [64].

**D. Electronic structure of spin 3/2 color centers in SiC polytypes**

A schematic representation of the crystal structure with color centers in 6H-SiC is presented in Fig. 13. The direction of the c-axis is shown, the staircase drawn with the heavy lines. In the bottom the orientation of the x, y, z axes are indicated. The {11-20} plane, contain the c-axis (z), x-axis and run parallel to the surface of the paper. The V2 and V1/V3 color centers as $V_{Si}^{-}$ - $V_C^0$ structures are indicated. Groups of C and Si nuclei indicated by $C_{NN}(1\text{-}4)_{Si}$ and $Si_{NNN}(1\text{-}9)_{Si}$ correspond to the C nearest neighbors and the Si next-nearest neighbors according the silicon vacancy $V_{Si}^{-}$. $Si_{NN}(1\text{-}4)_C$ indicate the Si nearest neighbors according the carbon vacancy $V_C^0$, note that for the V2 center (only) the $Si_{NN}(2\text{-}4)_C$ are simultaneously $Si_{NNN}(10\text{-}12)_{Si}$ that is, the Si next-nearest neighbors according the silicon vacancy $V_{Si}^{-}$. $C_{NNN}(n)_C$ correspond to the C next-nearest neighbors (NNN) according the carbon vacancy $V_C^0$.

The Roman numbers $III_{Si}$, $IV_{Si}$ $III_C$, $IV_C$ placed near the carbon and silicon atoms, respectively, correspond to the numbers indicating the shell number according the silicon vacancy $V_{Si}^-$ or the carbon vacancy $V_C^0$ (only part of the atoms are shown as an example).

The polytypes 4H-SiC, 6H-SiC and 15R-SiC show very similar V1, V2, V3 and V4 hyperfine structure and electron *g* tensors, in particular, the V2 parameters in all these polytypes are almost identical. These facts together suggest that these color centers exhibit the same microscopic structure in all the polytypes. ENDOR measurements yielded the important result that the HF interaction parameters can have opposite signs.

As a starting point for constructing a model of a spin color center with a spin 3/2, we consider HF interactions whose nature has been reliably established in previous EPR and ENDOR studies.

1. The HF interaction with the four carbon atoms surrounding the negatively charged silicon vacancy and located at the vertices of the tetrahedron, the so-called the nearest-neighbour $C_{NN}$ atoms. One of the carbon atoms being located at the apex, the bond of which with the silicon vacancy coincides with the c axis, and the other three carbon atoms are located in basal plane. The values of these HF interactions are presented in Table II.
2. The HF interaction with twelve silicon atoms located in the second coordination sphere relative to the silicon vacancy, the so-called the next nearest-neighbour $Si_{NNN}$ atoms (approximate values of the HF interactions obtained from the EPR results are given in Table II).
3. The color centers are defects with deep levels in the band gap; therefore, the wave functions of these centers are quite strongly localized in comparison with shallow donors and shallow acceptors (see Refs. [65-72]).

We consider the first four coordination spheres with respect to both silicon vacancies and carbon vacancies, which, in our opinion, are included in the structure of the spin center. Regarding the silicon vacancy $V_{Si}^-$, we have the following coordination spheres: (I) $C_{NNSi}$, (II) $Si_{NNNSi}$, (III) $C_{IIISi}$, and (IV) $Si_{IVSi}$. Relative to the carbon vacancy $V_C^0$, respectively: (I) $Si_{NNC}$, (II) $C_{NNNC}$, (III) $Si_{IIIC}$, and (IV) $C_{IVC}$ (the index Si or C is omitted when the symmetry core is mentioned in the text). The HF interaction constants that we attribute to these coordination spheres are given in Table II.

Let us consider polytype 6H-SiC, since for this polytype there is the most complete set of experimental data obtained by the ENDOR method (Table II).

There are several groups of lines, each of which is characterized by close values of hyperfine interactions. We will consider them under the assumption that the absolute values of these interactions decrease with increasing distance from the center with the maximum spin density, and this decrease should be significantly larger compared with the corresponding decrease in HF interactions for shallow donors or shallow acceptors. Important characteristics are also the degree of anisotropy of the HF interactions and the sign of the spin density. It is noteworthy that the form of the HF interaction anisotropy for silicon and carbon is similar for some values, i.e., they can be attributed to the same coordination spheres with respect to the centers of symmetry, for which a silicon vacancy or a carbon vacancy can be taken.

In works [65-72], the EPR and ENDOR methods were used to diagnose the spatial distribution of the wave functions of unpaired electrons for shallow nitrogen donors and shallow boron acceptors in various silicon carbide polytypes. It is important to note that nitrogen occupies the carbon position, while boron is in the silicon position, i.e., based on these studies, it was possible to study the spatial distribution of the spin density both with the center in the carbon site and in the silicon site, which is important since our spin center model includes silicon vacancy and carbon vacancy.

The HF interactions with $^{29}$Si located in NNN positions with positive HF interaction constants (negative spin density on the Si nucleus) around the silicon vacancy of the S = 3/2 center are almost isotropic, meanwhile, interactions with $^{13}$C located in the NN shell around the silicon vacancy are strongly anisotropic (see Table II). Some ENDOR signals for HF interaction with $^{29}$Si exhibit anisotropic dependence typical of the HF interactions between $^{13}$C located in the NN shell around $V_{Si}^-$ (see EPR spectrum in Fig. 3 and Fig. 8). These strongly anisotropic ENDOR signals, which correspond to negative constants of HF interaction with $^{29}$Si nuclei, i.e., a positive spin density on the Si nucleus, are shown in Figs. 2-4, 6-9. In Fig. 4, the HF interaction constants of these signals are designated as $A_1$ and $A_{2-4}$, while $A_1$ corresponds to the HF interaction with the nearest-neighbor $Si_{NN}$ atoms located along the c axis of the crystal for the tetrahedron around the carbon vacancy, while $A_{2-4}$ corresponds to the HF interaction with three $Si_{NN}$ atoms located in the basal plane of the crystal. Therefore, to explain such anisotropy, we need to identify Si atoms that have the same symmetry as the C atoms around a silicon vacancy $V_{Si}^-$. Such a configuration can be found only at the tetrahedron vertices around the carbon site. The position of the ENDOR lines labeled as $A_1$ and $A_{2-4}$ in Fig. 4 agrees well with the proposed configuration and reflects the HF interactions with axial ($Si_1$) and basal ($Si_{2–4}$) nuclear spins. The HF interactions with the negative $^{29}$Si HF interaction constant $Si_{NN}(I)_C$ are relatively large and describe well the anisotropy of the linewidth observed in the ESE detected EPR (Fig. 8 a, inset), that is, this anisotropy also manifests itself in the EPR spectra.

The observed HF interactions with the positive spin density on the $^{29}$Si nuclei can be explained if the spin density is located on four Si nuclei placed around nonparamagnetic neutral $V_C^0$. This implies that the spin center is formed by both non-paramagnetic $V_C^0$ and paramagnetic S = 3/2 $V_{Si}^-$. The spin density is caused by the electron spin polarization (similar to the core polarization for transition metals, which can be negative, e.g., for internal ns paired electrons or positive for external ns paired electrons [64]), and arises from an exchange interaction with S = 3/2 that leads to the partial decoupling of coupled covalent bonds of the $V_C^0$ site. The presence of the $V_C^0$ distorts the crystal lattice, which in turn lowers the symmetry of the silicon vacancy in regular defect-free environment ($V_{Si}^-$). Because the V2 center is characterized by the largest zero-field splitting we can conclude that $V_{Si}^-$ and $V_C^0$ are located closer to each other than in the case of V1, V3, V4 centers. The duplication of lines observed in the ENDOR spectra can be explained by the presence of two similar centers with slightly different parameters of the HF interactions.

Table II presents ligand hyperfine parameters for interaction with $^{13}$C-atoms and $^{29}$Si-atoms surrounding a spin color center of V1, V2, V3, V4 in three polytypes of SiC: 6H-, 15R- and 4H-SiC. The HF-tensor principal values *a* and *b* and the corresponding *s* and *p* spin densities of the unpaired electron connected to the spin centers in SiC with the nuclei surrounding the center. To consider the distribution of *s* and *p* character, the observed HF interactions were translate into spin density using the table of Morton and Preston [73]. The isotropic HF interaction (*a*) gives a measure of the s spin density (s), whereas the anisotropic hf interaction (*b*) is a measure for the p density (p).

The HF interaction with four $^{13}$C-atoms in the NN shell has axial symmetry along the bonding direction, therefore parameters are given parallel to the bond ($A_{\|}$) and perpendicular to the bond ($A_{\perp}$), from EPR data [56]. The HF interaction with the twelve $^{29}$Si-atoms in the NNN shell is almost isotropic (EPR data [56]).

Based on our assumption that the structure of the spin center includes both a silicon vacancy and a carbon vacancy, we will consider four nearest coordination spheres. Around $V_{Si}^-$– $C_{NN}$(I)[“+”, anisotropic] $Si_{NNN}$(II)[“-”, isotropic] C(III)[“+”, anisotropic] Si(IV)[“+”, isotropic]. Around $V_C^0$– $Si_{NN}$(I)[“+”, anisotropic] $C_{NNN}$(II)[“-”, isotropic] Si(III)[“-”, anisotropic] C(IV)[“-”, small anisotropy]. Plus “+”/ minus “-“ indicates the positive / negative spin density on the corresponding atom; isotropic / anisotropic characterizes the isotropic or anisotropic distribution of spin density on the corresponding atom. The anisotropic distribution is described by the constants *a* and *b*, which are estimated approximately using a set of hyperfine interactions defined for different orientations. Registration of complete orientation dependencies is a time-consuming task and was performed only for a few shells. For some shells, when the number of atoms is known, the total spin density is estimated (see Table II).

## IV. CONCLUSIONS

The polytypes 4H-SiC, 6H-SiC and 15R-SiC show very similar V1, V2, V3 and V4 hyperfine structure and electron *g* tensors, in particular, the V2 parameters in all these polytypes are almost identical. These facts together suggest that these color centers exhibit the same microscopic structure in all polytypes. Based on EPR and ENDOR studies, the signs of fine structure splitting for different spin centers were determined, the structure of spin energy levels in magnetic fields, including the zero magnetic field, was uniquely established. Populations of spin levels created by optical pumping at different temperatures, including room temperatures and above, were found. ENDOR measurements yielded the important result that the HF interaction parameters can have opposite signs. The observed hyperfine interactions were shown to be a strong evidence in favor of the model as a negatively charged paramagnetic silicon vacancy that is noncovalently bonded to the non-paramagnetic neutral carbon vacancy, located on the adjacent site along the SiC symmetry c-axis, i.e., the $V_{Si}^{-}$ - $V_C^0$ model with S = 3/2. We emphasize the most important reasons for choosing the proposed model.

It was observed the same anisotropy of $^{13}$C HF interactions for carbon $C_{NN}$ found from the EPR spectra and $^{29}$Si HF interactions for $Si_{NN}$ found from the ENDOR spectra, reflecting the HF interaction with the four nearest carbon atoms surrounding the silicon vacancy, on the one hand, and the HF interaction with the four nearest silicon atoms surrounding the carbon vacancy, on the other hand. These results are evidence of core of symmetry in the form of a silicon vacancy or a carbon vacancy. In this case, the spin density on four $C_{NN}$ nuclei around a negatively charged silicon vacancy $V_{Si}^{-}$ [$C_{NN}(V_{Si}^{-})_{1\text{-}4}$, Table II] is due to three unpaired silicon vacancy electrons, S=3/2 system, and the spin density on four $Si_{NN}$ nuclei near a neutral carbon vacancy $V_C^0$ [$Si_{NN}(V_C^0)_{1\text{-}4}$, Table II] is associated with a core polarization mechanism leading to incomplete pairing of closed valence shells.

The presence of identical orientational dependences of the $^{29}$Si ENDOR signals for silicon near the Larmor frequency of $^{29}$Si [$Si_{III}(V_C^0)$, Table II], on the one hand, and the ENDOR signals for carbon $^{13}$C near the Larmor frequency of $^{13}$C [$C_{III}(V_{Si}^{-})$, Table II], on the other hand, indicate the presence of silicon coordination sphere relative to the silicon vacancy for $^{29}$Si HF interactions and identical carbon coordination sphere relative to the carbon vacancy for $^{13}$C HF interactions. These results are additional evidence of core of symmetry in the form of a silicon vacancy or a carbon vacancy.

A fundamental difference in the EPR, ENDOR, photoluminescence and ODMR spectra between axial spin 3/2 centers $V_{Si}^{-}$ - $V_C^0$, on the one hand, and a negatively charged silicon vacancy $V_{Si}^{-}$ in a regular defect-free environment, on the other, has been established. In the latter case, no photoluminescence and optically induced alignment of spin level populations were observed, and no ODMR was detected.

## ACKNOWLEDGMENTS

This work was supported by the Russian Science Foundation (Project No. 20-12-00216).

## References

[1] G. Watkins, in *Deep Centers in Semiconductors*, edited by S.T. Pantelides (Gordon and Breach, New York, 1986), p. 147.

[2] A. Gruber, A. Drabenstedt, C. Tietz, L. Fleury, J. Wrachtrup, C. von Borczyskowski, Scanning confocal optical microscopy and magnetic resonance on single defect centers. Science **276**, 2012-2014 (1997).

[3] F. Jelezko, T. Gaebel, I. Popa, A. Gruber, J. Wrachtrup, Observation of coherent oscillations in a single electron spin. Phys. Rev. Lett. **92**, 076401 (2004).

[4]. R. Hanson, F. Mendoza, R. J. Epstein, D. D. Awschalom, Phys. Rev. Lett. **97**, 087601 (2006).

[5] F. Jelezko, J. Wrachtrup, Single defect centers in diamond: A review. Phys. Status Solidi A **203**, 3207-3225 (2006).

[6] D. D. Awschalom, M. E. Flatté, Challenges for semiconductor spintronics. Nature Phys. **3**, 153-159 (2007).

[7] R. Hanson, D. D. Awschalom, Coherent manipulation of single spins in semiconductors. Nature **453**, 1043-1049 (2008).

[8] M. Koenraad, M. E. Flatté, Single dopants in semiconductors. Nature Mater. **10**, 91-100 (2011).

[9] P. G. Baranov, I. V. Il'in, E. N. Mokhov, M. V. Muzafarova, S. B. Orlinskii, J. Schmidt, EPR identification of the triplet ground state and photoinduced population inversion for a Si-C divacancy in silicon carbide. JETP Lett. **82**, 441-443 (2005).

[10] P. G. Baranov, A. P. Bundakova, I. V. Borovykh, S. B. Orlinskii, R. Zondervan, J. Schmidt, Spin polarization induced by optical and microwave resonance radiation in a Si vacancy in SiC: A promising subject for the spectroscopy of single defects. JETP Lett. **86**, 202-206 (2007).

[11] J. R. Weber, W. F. Koehl, J. B. Varley, A. Janotti, B. B. Buckley, C. G. Van de Walle, D. D. Awschalom, Quantum computing with defects. Proc. Natl. Acad. Sci. USA **107**, 8513-8518 (2010).

[12] D. DiVincenzo, Quantum bits: Better than excellent. Nature Mater. **9**, 468-469 (2010).

[13] P. G. Baranov, A. P. Bundakova, A. A. Soltamova, S. B. Orlinskii, I. V. Borovykh, R. Zondervan, R. Verberk, J. Schmidt, Silicon vacancy in SiC as a promising quantum system for single defect and single-photon spectroscopy. Phys. Rev. B **83**, 125203 (2011).

[14] A. G. Smart, Silicon carbide defects hold promise for device-friendly qubits. Phys. Today **65**, 10-11 (2012).

[15] A. I. Veinger, V. A. Il'in, Yu. M. Tairov, V. F. Tsvetkov, Investigation of thermal defects in silicon carbide by the ESR method. Sov. Phys. Semicond. **13**, 1385 (1979).

[16] V. S. Vainer, V. A. Il'in, Electron spin resonance of exchange-coupled vacancy pairs in hexagonal silicon carbide. Sov. Phys. Solid State **23**, 2126-2133 (1981).

[17] H. J. von Bardeleben, J. L. Cantin, I. Vickridge, G. Battistig, Proton-implantation-induced defects in n-type 6H- and 4H−SiC: An electron paramagnetic resonance study. Phys. Rev. B **62**, 10126-10134 (2000).

[18] H. J. von Bardeleben, J. L. Cantin, L. Henry, M. Barthe, Vacancy defects in p-type 6H−SiC created by low-energy electron irradiation. Phys. Rev. B **62**, 10841-10846 (2000).

[19] M. Wagner, B. Magnusson, W. M. Chen, E. Janzen, E. Sörman, C. Hallin, J. L. Lindström, Electronic structure of the neutral silicon vacancy in 4H and 6H SiC. Phys. Rev. B **62**, 16555-16560 (2000).

[20] N. Mizuochi, S. Yamasaki, H. Takizawa, N. Morishita, T. Ohshima, H. Itoh, J. Isoya, Continuous-wave and pulsed EPR study of the negatively charged silicon vacancy with S=3/2 and C3v symmetry in n-type 4H-SiC. Phys. Rev. B **66**, 235202 (2002).

[21] W. E. Carlos, N. Y. Garces, E. R. Glaser, M. A. Fanton, Annealing of multivacancy defects in 4H−SiC. Phys. Rev. B **74**, 235201 (2006).

[22] S. B. Orlinski, J. Schmidt, E. N. Mokhov, P. G. Baranov, Silicon and carbon vacancies in neutron-irradiated SiC: A high-field electron paramagnetic resonance study. Phys. Rev. B **67**, 125207 (2003).

[23] W. F. Koehl, B. B. Buckley, F. J. Heremans, G. Calusine, D. D. Awschalom, Room temperature coherent control of defect spin qubits in silicon carbide. Nature **479**, 84-87 (2011).

[24] V. A. Soltamov, A. A. Soltamova, P. G. Baranov, I. I. Proskuryakov, Room temperature coherent spin alignment of silicon vacancies in 4H- and 6H-SiC. Phys. Rev. Lett. **108**, 226402 (2012).

[25] D. Riedel, F. Fuchs, H. Kraus, S. Vath, A. Sperlich, V. Dyakonov, A. A. Soltamova, P. G. Baranov, V. A. Ilyin, G. V. Astakhov, Resonant addressing and manipulation of silicon vacancy qubits in silicon carbide. Phys. Rev. Lett. **109**, 226402 (2012).

[26] F. Fuchs, V. A. Soltamov, S. Vath, P. G. Baranov, E. N. Mokhov, G. V. Astakhov, V. Dyakonov, Silicon carbide light-emitting diode as a prospective room temperature source for single photons. Sci. Rep. **3**, 1637 (2013).

[27] S. Castelletto, B. C. Johnson, A. Boretti, Quantum effects in silicon carbide hold promise for novel integrated devices and sensors. Adv. Opt. Mater. **1**, 609-625 (2013).

[28] A. L. Falk, B. B. Buckley, G. Calusine, W. F. Koehl, V. V. Dobrovitski, A. Politi, C. A. Zorman, P. X. L. Feng, D. D. Awschalom, Polytype control of spin qubits in silicon carbide. Nat. Commun. **4**, 1819 (2013).

[29] S. Castelletto, B. C. Johnson, V. Ivady, N. Stavrias, T. Umeda, A. Gali, T. Ohshima, A silicon carbide room-temperature single-photon source. Nature Mater. **13**, 151-156 (2013).

[30] T. C. Hain, F. Fuchs, V. A. Soltamov, P. G. Baranov, G. V. Astakhov, T. Hertel, V. Dyakonov, Excitation and recombination dynamics of vacancy-related spin centers in silicon carbide. J. Appl. Phys. **115**, 133508 (2014).

[31] S. Castelletto, B. C. Johnson, C. Zachreson, D. Beke, I. Balogh, T. Ohshima, I. Aharonovich, A. Gali, Room temperature quantum emission from cubic silicon carbide nanoparticles. ACS Nano **8**, 7938-7947 (2014).

[32] A. Muzha, F. Fuchs, N. V. Tarakina, D. Simin, M. Trupke, V. A. Soltamov, E. N. Mokhov, P. G. Baranov, V. Dyakonov, A. Krueger, G. V. Astakhov, Room-temperature near-infrared silicon carbide nanocrystalline emitters based on optically aligned spin defects. Appl. Phys. Lett. **105**, 243112 (2014).

[33] H. Kraus, V. A. Soltamov, D. Riedel, S. Vath, F. Fuchs, A. Sperlich, P. G. Baranov, V. Dyakonov, G. V. Astakhov, Room-temperature quantum microwave emitters based on spin defects in silicon carbide. Nature Phys. **10**, 157-162 (2014).

[34] G. Calusine, A. Politi, D. D. Awschalom, Silicon carbide photonic crystal cavities with integrated color centers. Appl. Phys. Lett. **105**, 011123 (2014).

[35] P. V. Klimov, A. L. Falk, B. B. Buckley, D. D. Awschalom, Electrically driven spin resonance in silicon carbide color centers. Phys.Rev. Lett. **112**, 087601 (2014).

[36] A. L. Falk, P. V. Klimov, B. B. Buckley, V. Ivady, I. A. Abrikosov, G. Calusine, W. F. Koehl, A. Gali, D. D. Awschalom, Electrically and mechanically tunable electron spins in silicon carbide color centers. Phys. Rev. Lett. **112**, 187601 (2014).

[37] H. Kraus, V. A. Soltamov, F. Fuchs, D. Simin, A. Sperlich, P. G. Baranov, G. V. Astakhov, V. Dyakonov, Magnetic field and temperature sensing with atomic-scale spin defects in silicon carbide. Sci.Rep. **4**, 5303 (2014).

[38] L.-P. Yang, C. Burk, M. Widmann, S.-Y. Lee, J. Wrachtrup, N. Zhao, Electron spin decoherence in silicon carbide nuclear spin bath. Phys. Rev. B **90**, 241203 (2014).

[39] V. A., Soltamov, B. V. Yavkin, D. O. Tolmachev, R. A. Babunts, A. G. Badalyan, V. Yu. Davydov, E. N. Mokhov, I. I. Proskuryakov, S. B. Orlinskii, P. G. Baranov, Optically addressable silicon vacancy-related spin centers in rhombic silicon carbide with high breakdown characteristics and ENDOR evidence of their structure. Phys. Rev. Lett. **115**, 247602 (2015).

[40] O. V. Zwier, D. O'Shea, A. R. Onur, C. H. van der Wal, All-optical coherent population trapping with defect spin ensembles in silicon carbide. Sci. Rep. **5**, 10931 (2015).

[41] A. L. Falk, P. V. Klimov, V. Ivady, K. Szasz, D. J. Christle, W. F. Koehl, A. Gali, D. D. Awschalom, Optical polarization of nuclear spins in silicon carbide. Phys. Rev. Lett. **114**, 247603 (2015).

[42] S. G. Carter, Ö. O. Soykal, P. Dev, S. E. Economou, E. R. Glaser, Spin coherence and echo modulation of the silicon vacancy in 4H-SiC at room temperature. Phys. Rev. B **92**, 161202 (2015)

[43] D. Simin, F. Fuchs, H. Kraus, A. Sperlich, P. G. Baranov, G. V. Astakhov, V. Dyakonov, High-precision angle-resolved magnetometry with uniaxial quantum centers in silicon carbide. Phys. Rev. Applied **4**, 014009 (2015).

[44] S.-Y. Lee, M. Niethammer, J. Wrachtrup, Vector magnetometry based on S=3/2 electronic spins. Phys. Rev. B **92**, 115201 (2015).

[45] D. J. Christle, A. L. Falk, P. Andrich, P. V. Klimov, J. Ul Hassan, N. T. Son, E. Janzen, T. Ohshima, D. D. Awschalom, Isolated electron spins in silicon carbide with millisecond coherence times. Nature Mater. **14**, 160-163 (2015).

[46] M. Widmann, S.-Y. Lee, T. Rendler, N. T. Son, H. Fedder, S. Paik, L.-P. Yang, N. Zhao, S. Yang, I. Booker, A. Denisenko, M. Jamali, S. A. Momenzadeh, I. Gerhardt, T. Ohshima, A. Gali, E. Janzen, J. Wrachtrup, Coherent control of single spins in silicon carbide at room temperature. Nature Mater. **14**, 164-168 (2015).

[47] F. Fuchs, B. Stender, M. Trupke, D. Simin, J. Paum, V. Dyakonov, G. V. Astakhov, Engineering near infrared single-photon emitters with optically active spins in ultrapure silicon carbide. Nat. Commun. **6**, 7578 (2015).

[48] A. Lohrmann, N. Iwamoto, Z. Bodrog, S. Castelletto, T. Ohshima, T. J. Karle, A. Gali, S. Prawer, J. C. McCallum, B. C. Johnson, Single-photon emitting diode in silicon carbide. Nat.Commun. **6**, 7783 (2015).

[49] T. Wimbauer, B.K. Meyer, A. Hofstaetter, A. Scharmann, and H. Overhof, Physical Review B **56**, 7384 (1997).

[50] H. J. Von Bardeleben, J. L. Cantin, E. Rauls, and U. Gerstmann, Physical Review B **92**, 064104 (2015).

[51] S. A. Zargaleh, B. Eble, S. Hameau, J. L. Cantin, L. Legrand, M. Bernard, F. Margaillan, J. S. Lauret, J. F. Roch, H. J. Von Bardeleben, E. Rauls, U. Gerstmann, and F. Treussart, Physical Review B **94**, 060102 (2016).

[52] H. J. Von Bardeleben, J. L. Cantin, A. Csóré, A. Gali, E. Rauls, and U. Gerstmann, Physical Review B 94, 121202 (2016).

[53] A. Csóré, H. J. Von Bardeleben, J. L. Cantin, and A. Gali, Physical Review B **96**, 085204 (2017).

[54] Yu. A. Vodakov, E. N. Mokhov, M. G. Ramm, and A. D. Roenkov, Krist. Tech. **5**, 729 (1979)]

[55] V.G. Grachev, Zh. Eksp. Teor. Phys. 92, 1834 (1987) [Sov. Phys. JETP **65**, 7029 (1987)].

[56] Mt. Wagner, N. Q. Thinh, N. T. Son, W. M. Chen, and E. Janzén, P. G. Baranov, E. N. Mokhov, C. Hallin, J. L. Lindström, Ligand hyperfine interaction at the neutral silicon vacancy in 4H- and 6H-SiC, Physical Review B **66**, 155214 (2002).

[57] T. Biktagirov, W. G. Schmidt, U. Gerstmann, B. Yavkin, S. Orlinskii, P. Baranov, V. Dyakonov, V. Soltamov, Polytypism driven zero-field splitting of silicon vacancies in 6H-SiC, Physical Review B **98**, 195204 (2018).

[58] J.-M. Spaeth, J. R. Niklas, and R. H. Bartram, *Structural Analysis of Point Defects in Solids* (Springer-Verlag, Berlin, Heidelberg, 1992), Chap. 5, p. 152.

[59] P. G. Baranov, H.-J. von Bardeleben, F. Jelezko, J. Wrachtrup, *Magnetic Resonance of Semiconductors and Their Nanostructures: Basic and Advanced Applications* (Springer Series in Materials Science, Volume 253, Springer-Verlag GmbH Austria, 2017) Chap. 6.

[60] W. B. Mims, in *Electron Paramagnetic Resonance*, edited by S. Geshwind (Plenum, New York, 1972), p. 344.

[61] C. Gemperle, A. Schweiger, Chem. Rev. **91**, 1481 (1991).

[62] G. Jeschke, A. Schweiger, Chem. Phys. Lett. **246**, 431 (1995).

[63] A. Abragam, B. Bleaney: *Electron Paramagnetic Resonance of Transition Ions*. Clarendon Press, Oxford, 1970, p. 702.

[64] C. J. Cochrane, J. Blacksberg, M. A. Anders, P. M. Lenahan, Vectorized magnetometer for space applications using electrical readout of atomic scale defects in silicon carbide, Scientific Reports **6**, 37077 (2016).

[65] T. Matsumoto, O. G Poluektov, J. Schmidt, E. N. Mokhov, P. G. Baranov, Electronic structure of the shallow boron acceptor in 6H-SiC:mA pulsed EPR/ENDOR study at 95 GHz. Phys. Rev. B **55**, 2219-2229 (1997).

[66] A. van Duijn-Arnold, R. Zondervan, J. Schmidt, P. G. Baranov, E. N. Mokhov, Electronic structure of the N donor centre in 4H-SiC and 6H-SiC, Phys. Rev. B **64**, 085206 (2001).

[67] N. T. Son, E. Janzén, J. Isoya, S. Yamasaki, Hyperfine interaction of the nitrogen donor in 4H−SiC. Phys. Rev. B **70**, 193207 (2004).

[68] D. V. Savchenko, E. N. Kalabukhova, V. S. Kiselev, J. Hoentsch, A. Poppl, Spin-coupling and hyperfine interaction of the nitrogen donors in 6H-SiC. Phys. Status Solidi B **246**, 1908-1914 (2009).

[69] D. V. Savchenko, E. N. Kalabukhova, A. Poppl, B. D. Shanina, Electronic structure of the nitrogen donors in 6H SiC as studied by pulsed ENDOR and TRIPLE ENDOR spectroscopy. Phys. Status Solidi B **249**, 2167-2178 (2012).

[70] P. G. Baranov, B. Ya. Ber, O. N. Godisov, I. V. Il'in, A. N. Ionov, E. N. Mokhov, M. V. Muzafarova, A. K. Kaliteevskii, M. A. Kaliteevskii, P. S. Kop'ev, Probing of the shallow donor and acceptor wave functions in silicon carbide and silicon through an EPR study of crystals with a modified isotopic composition. Phys. Solid State **47**, 2219–2232 (2005).

[71] S. Greulich-Weber, M. Feege, J.-M. Spaeth, E. N. Kalabukhova, S. N. Lukin, E. N. Mokhov, On the microscopic structures of shallow donors in 6H SiC: studies with EPR and ENDOR. Solid State Commun. **93**, 393-397 (1995).

[72] N. T. Son, J. Isoya, T. Umeda, I. G. Ivanov, A. Henry, T. Ohshima, E. Janzén, EPR and ENDOR studies of shallow donors in SiC, Appl Magn Reson **39**, 49–85 (2010).

[73] J. R. Morton and K. F. Preston, Atomic parameters for paramagnetic resonance data, J. Magn. Res. **30**, 377 (1978).

## Figure captions

FIG. 1. ESE-detected EPR spectra of V2 centers observed at W-band in 4H-SiC (a), 6H-SiC (b) crystals measured at low temperature (1.2 K) and high temperature (190 K and 50 K) at two orientations between the magnetic field B and the crystal c-axis without light excitation (light off) with Boltzmann distribution. (c) EPR spectra of V2 and V1/V3 centers observed at X-band in 6H-SiC in orientation B || c at low temperature of 5 K and high temperature of 18 K without light, in addition, the EPR spectra under light excitation are shown at 5 K and 60 K. (d) ESE-detected EPR spectra of V2 centers observed at W-band in 15R-SiC at three temperatures (50, 10, 8 K) for the orientation close to B || c without light, dashed line shows EPR spectra under 780 nm light excitation at 50 K.

FIG. 2. ESE detected ENDOR spectra at W-band of the V2 and V1/V3 centers in 6H-SiC, B || c, recorded with a scan in a wide frequency range, for the high-field (hf) transitions indicated in optically induced ESE detected EPR spectra shown at the right. The ENDOR lines with the strongest HF interactions (in absolute value) are marked. There is a mirror reflection of the HF interaction lines with different coordination spheres of silicon relative to the Larmor frequency of $^{29}$Si, which unambiguously indicates opposite signs of fine structure splitting. The insert shows one EPR line of V2 center in 6H-SiC crystal with a natural isotope silicon content and with a modified isotopic composition of silicon (0.7% of $^{29}$Si isotope).

FIG. 3. W-band angular dependence ESE-detected EPR (a) and corresponding ESE-detected ENDOR (b) spectra of the optically aligned V2 color centers in single 6H-SiC crystal. (c) Expanded-scale ESE-detected ENDOR spectrum from (b). The EPR transitions for the low-field (lf) and high-field (hf) magnetic fields are indicated in optically induced ESE spectra. The dotted lines in the (b, c) correspond to the signals recorded in the additional 6H-SiC sample. The angular dependencies of the ENDOR lines are highlighted in gray. (b) The signals at $f_L \pm 1/2|A_{SiNNN}|$ and $f_L \pm 3/2|A_{SiNNN}|$ correspond to HF interactions with the next-nearest-neighbors (NNN) Si atoms (with $^{29}$Si nuclei) with respect to the negatively charged silicon vacancy $V_{Si}^-$. The light-induced inverse population of the spin sublevels of V2 centers is shown in the top inset. Bottom inset displays a resolved in the EPR anisotropic hyperfine structure for the nearest-neighbors (NN) carbon atoms (with $^{13}$C nuclei) with respect to negatively charged silicon vacancy $V_{Si}^-$; and almost isotropic hyperfine structure for the next-nearest-neighbors (NNN) Si atoms (with $^{29}$Si nuclei) with respect to the negatively charged silicon vacancy $V_{Si}^-$. (c) The signals at $f_L \pm 1/2|A_{SiNN}|$ and $f_L \pm 3/2|A_{SiNN}|$ indicate ENDOR lines corresponding to the HF interactions with the nearest-neighbors (NN) Si atoms (with $^{29}$Si nuclei)

with respect to the neutral carbon vacancy $V_C^0$. The spectra for the perpendicular orientation (B⊥c) is given for demonstrating a small orientation dependence for some ENDOR signals.

FIG. 4. The manifestation of the blind spot effect of Mims-type ENDOR into experiments and affected the part of spectrum, preventing ENDOR lines at frequencies 29.6 MHz and 31.8 MHz from being observed in two upper plots. Notation of HF interaction $A_1$ was introduced for the ENDOR lines of the nearest-neighbor $Si_{NN}$ atoms around a carbon vacancy $V_C^0$ located along the *c* axis of the tetrahedron around the carbon vacancy, then $A_{2-4}$ correspond to three $Si_{NN}$ atoms in the basal plane. (dashed line) Reconstruction of the ENDOR spectra from several samples recorded at different values of τ.

FIG. 5. Angular dependence of W-band ESE-detected ENDOR for V2 color centers in single 6H-SiC crystal. Hyperfine interactions with the surrounding silicon nuclei $^{29}Si$ (the left-hand side) and carbon $^{13}C$ (right-hand part) are presented. The angular dependencies of the ENDOR lines are highlighted in gray and by dashed lines.

FIG. 6. W-band angular dependence ESE-detected EPR (a) and ESE-detected ENDOR (b) spectra of the optically aligned V1/V3 color centers in single 6H-SiC crystal. (c) Expanded-scale ESE-detected ENDOR spectrum from (b). The EPR transitions for the low-field (lf) and high-field (hf) magnetic fields are indicated in optically induced ESE spectra. The signals at $f_L \pm 1/2|A_{SiNNN}|$ and $f_L \pm 3/2|A_{SiNNN}|$ (b) correspond to HF interactions with the next-nearest-neighbors Si atoms (with $^{29}Si$ nuclei) with respect to the negatively charged silicon vacancy $V_{Si}^-$. (top inset) The light-induced inverse population of the spin sublevels of V1/V3 centers (D<0). (bottom inset) expanded-scale ESE-detected ENDOR signals with carbon atoms (with $^{13}C$ nuclei). The signals at $f_L \pm 1/2|A_{SiNN}|$ and $f_L \pm 3/2|A_{SiNN}|$ (c) indicate ENDOR lines corresponding to the HF interactions with the nearest-neighbors Si atoms (with $^{29}Si$ nuclei) with respect to the neutral carbon vacancy $V_C^0$.

FIG. 7. ESE detected ENDOR spectra at W-band of the V2 centers in 15R-SiC, B~|| c for the low-field (lf) and high-field (hf) transitions indicated in optically induced ESE detected EPR spectra shown at the right. Triangles denote the ENDOR lines with the strongest HF interactions (in absolute value) with negative HF constants.

FIG. 8. W-band angular dependence ESE-detected EPR (a) and corresponding ESE-detected ENDOR (b) spectra of the V2 color centers in single 15R-SiC crystal. The angular dependencies of the ENDOR lines for $Si_{NN}$ atoms are highlighted in gray.

FIG. 9. ESE detected ENDOR spectra at W-band of the V2, V3 and V4 color centers in 15R-SiC measured at orientation $\theta \approx 10^0$ for the low-field (lf) and high-field (hf) transitions indicated in optically induced ESE spectra shown at the right.

FIG. 10. ESE detected ENDOR spectra at W-band of the V2 color centers in 4H-SiC measured at orientation θ ≈$0^0$ for the low-field (lf) and high-field (hf) transitions indicated in optically induced ESE spectra shown at the right. For comparison, the $^{29}$Si ENDOR spectra of the V4 centers in polytype 15R-SiC and V1/V3 centers in 6H-SiC and $^{13}$C ENDOR spectra of the V2 center in 6H-SiC are shown. The inset shows the level system for the V2 center in 4H-SiC.

FIG. 11. A cw (a) and ESE (b) W-band spectra (94.9 GHz) of the $V_{Si}^-$ vacancy observed at 300 K (a) and 1.2 K (b) in n-irradiated 4H-SiC (dose of $10^{18}$ $cm^{-2}$) for several orientations of the magnetic field with respect to the c axis including the orientations parallel (θ=0°) and perpendicular (θ=90°) to the c axis. The central line and two HF satellites are shown for the h and k sites. The intensity ratio of the central line to that of the satellites corresponds to the interaction with 12 silicon atoms of the second shell. The results of simulation of the EPR (a) and ESE (b) spectra are presented.

FIG. 12. ESE detected ENDOR spectra at W-band of the negatively charged silicon vacancy in regular environment, $V_{Si}^-$ center, in 15R-SiC (for two orientations), which are characterized by zero-field splitting D = 0. Inset shows the ESE detected EPR for $V_{Si}^-$ centers in 15R-SiC.

FIG. 13. A schematic representation of the crystal structure with color centers in 6H-SiC. The direction of the c-axis is shown, the staircase drawn with the heavy lines. In the bottom the orientation of the x, y, z axes are indicated. The {11-20} plane, contain the c-axis (z), x-axis and run parallel to the surface of the paper. The V2 and V1(V3) color centers as $V_{Si}^-$ - $V_C^0$ structures are indicated. Groups of C and Si nuclei indicated by $C_{NN}(1\text{-}4)_{Si}$ and $Si_{NNN}(1\text{-}9)_{Si}$ correspond to the C nearest neighbors (NN) and the Si next-nearest neighbors (NNN) according the silicon vacancy $V_{Si}^-$. $Si_{NN}(1\text{-}4)_C$ indicate the Si nearest neighbors according the carbon vacancy $V_C^0$, note that for the V2 center (only) the $Si_{NN}(2\text{-}4)_C$ are simultaneously $Si_{NNN}(10\text{-}12)_{Si}$ that is, the Si next-nearest neighbors according the silicon vacancy $V_{Si}^-$. $C_{NNN}(n)_C$ correspond to the C next-nearest neighbors (NNN) according the carbon vacancy $V_C^0$. The Roman numbers $III_{Si}$, $IV_{Si}$ $III_C$, $IV_C$ placed near the carbon and silicon atoms, respectively, correspond to the numbers indicating the shell number according the silicon vacancy $V_{Si}^-$ or the carbon vacancy $V_C^0$ (only part of the atoms are shown as an example). Distance from $V_{Si}^-$ to $C_{NN}(1\text{-}4)$ is of 1.88-1.89 Å, to $Si_{NNN}(1\text{-}12)$ is of 3.07-3.09 Å, to CIII(n) - of 4.41, 4.75 Å. Distances from $V_C^0$ to $Si_{NN}(1\text{-}4)$ is of 1.88-1.89 Å.

FIG. 14. Optically-induced level spin-alignment at room temperature (RT), B=0 (for V1 and V3 centers in 6H-SiC optically-induced level spin-alignment is also shown for 30 K, B=0).

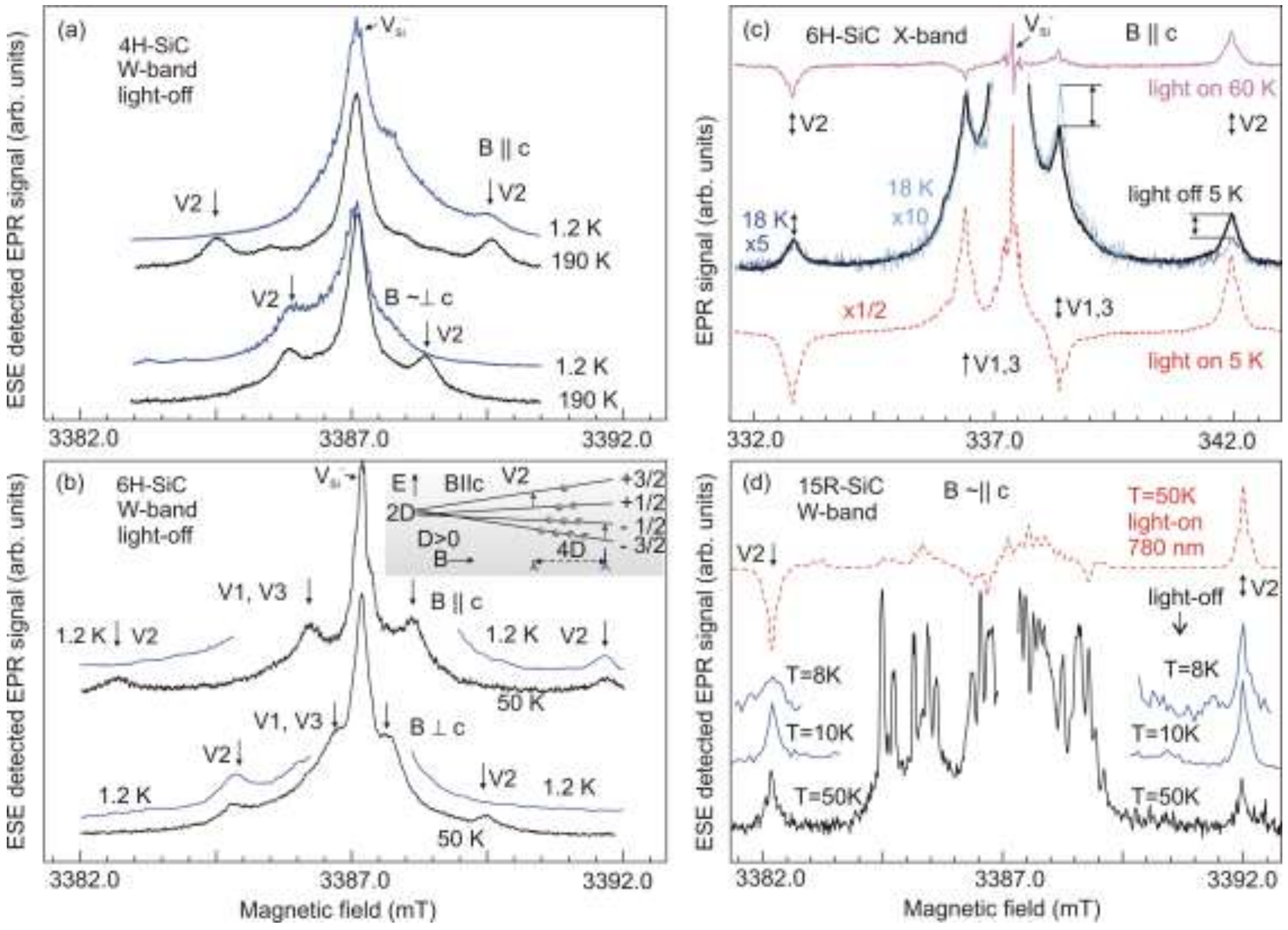

FIG. 1.

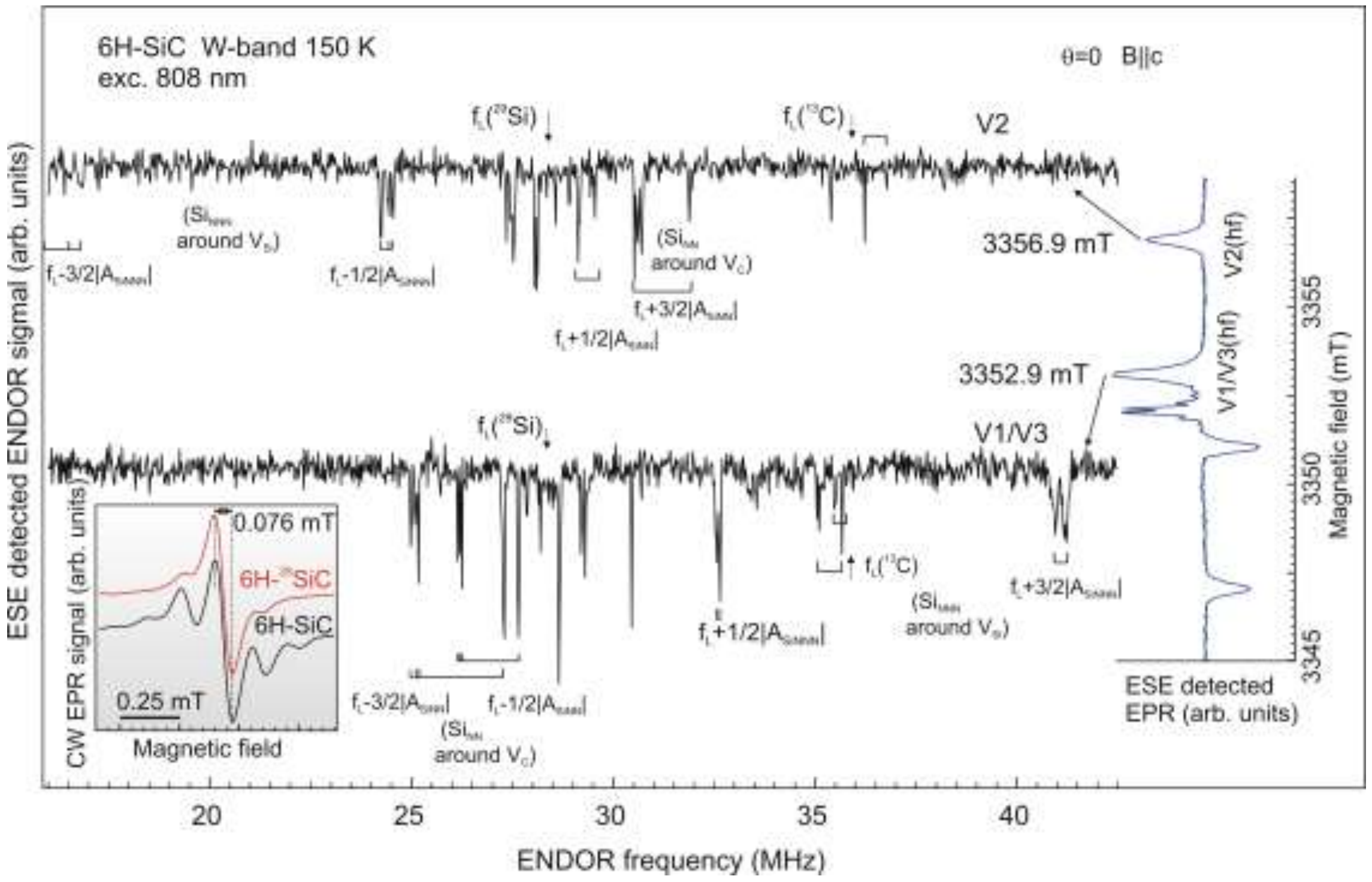

FIG. 2.

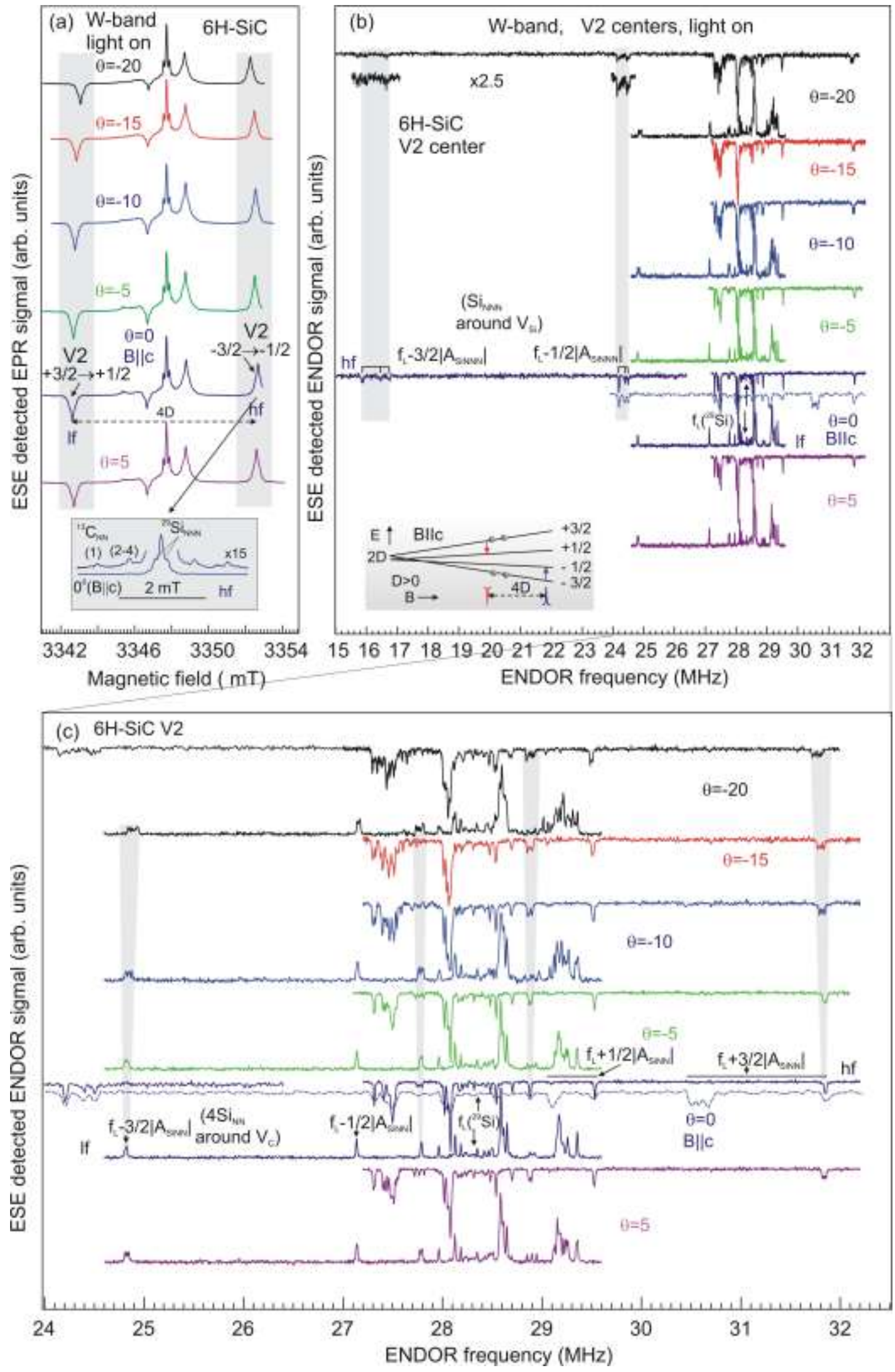
(a) W-band light on
6H-SiC
θ=-20
θ=-15
θ=-10
θ=-5
θ=0 B||c
θ=5
V2 +3/2→+1/2
V2 -3/2→-1/2
4D
lf
hf
2 mT
x15
3342 3346 3350 3354
Magnetic field ( mT)
ESE detected EPR signal (arb. units)
(b) W-band, V2 centers, light on
x2.5
6H-SiC V2 center
ESE detected ENDOR signal (arb. units)
ENDOR frequency (MHz)
(c) 6H-SiC V2
ESE detected ENDOR signal (arb. units)
ENDOR frequency (MHz)

FIG. 3.

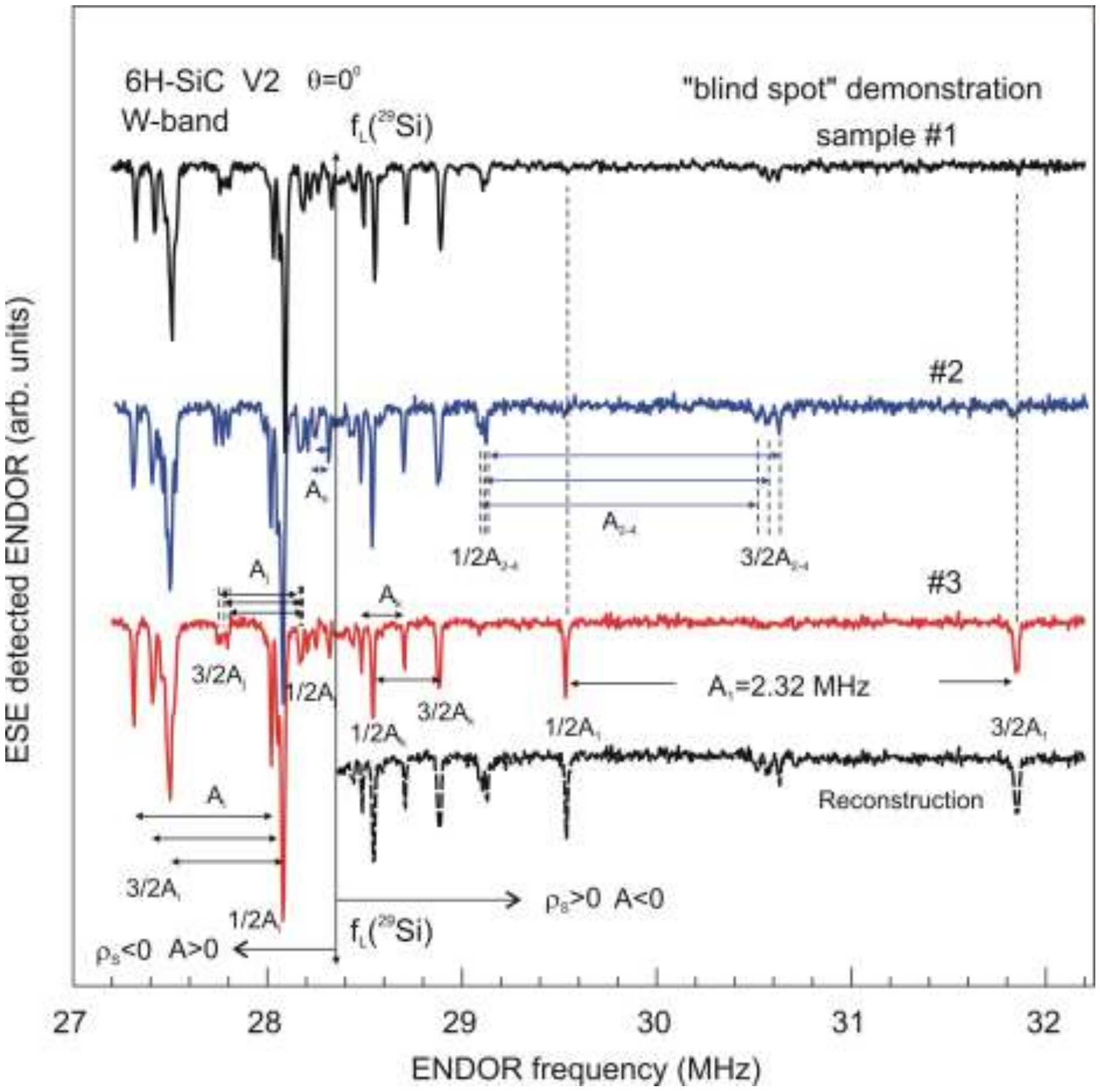

FIG. 4.

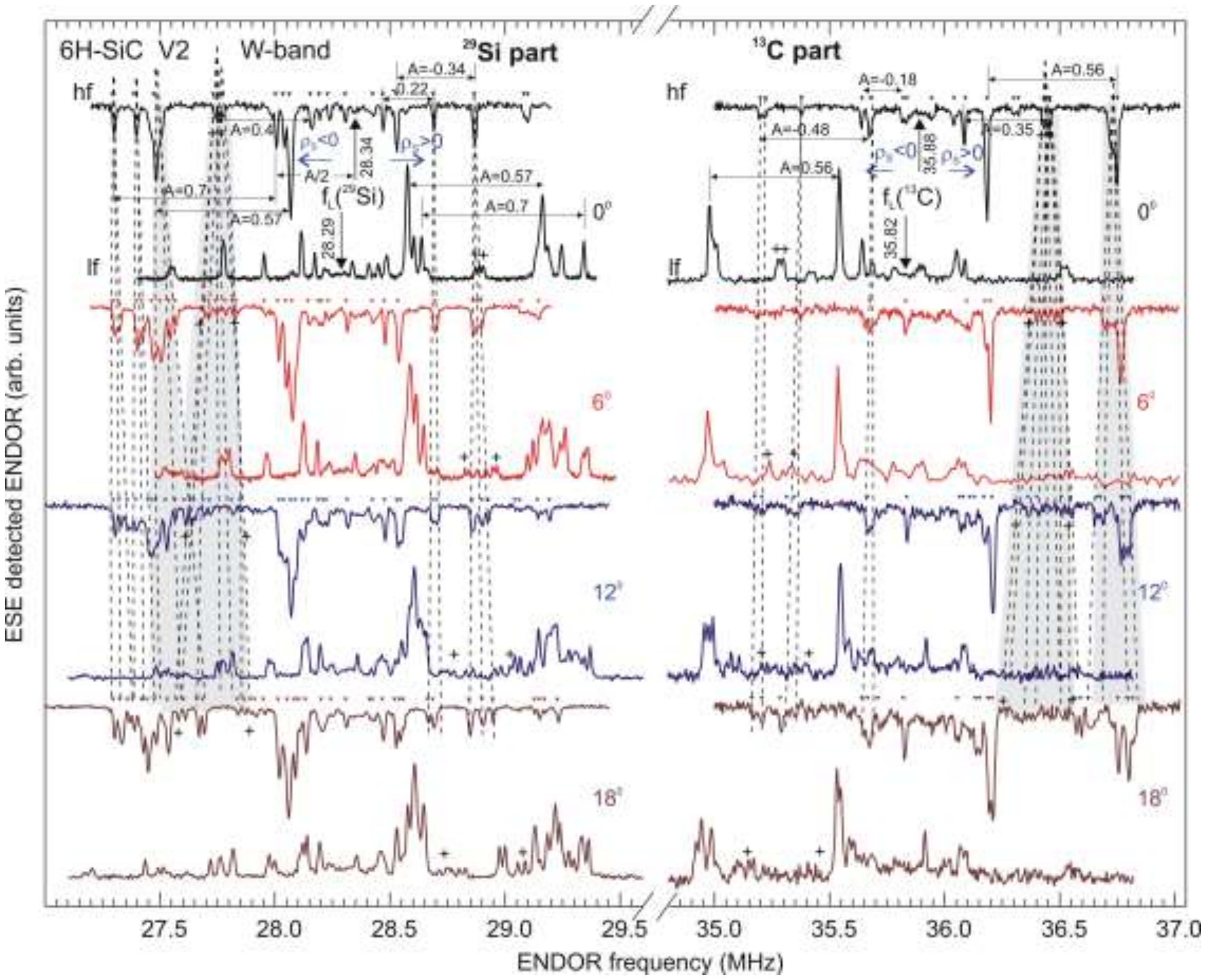

FIG. 5.

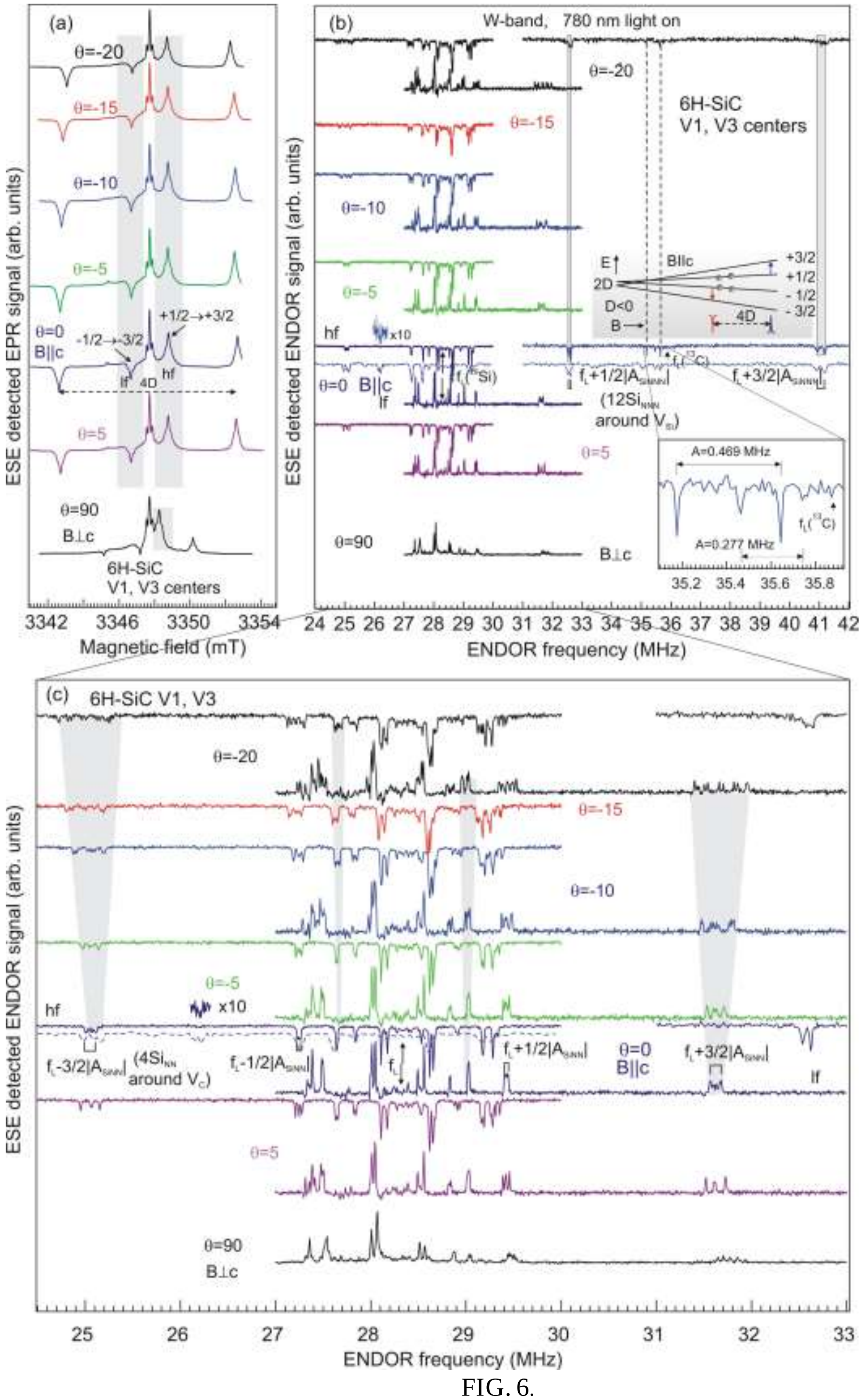

FIG. 6.

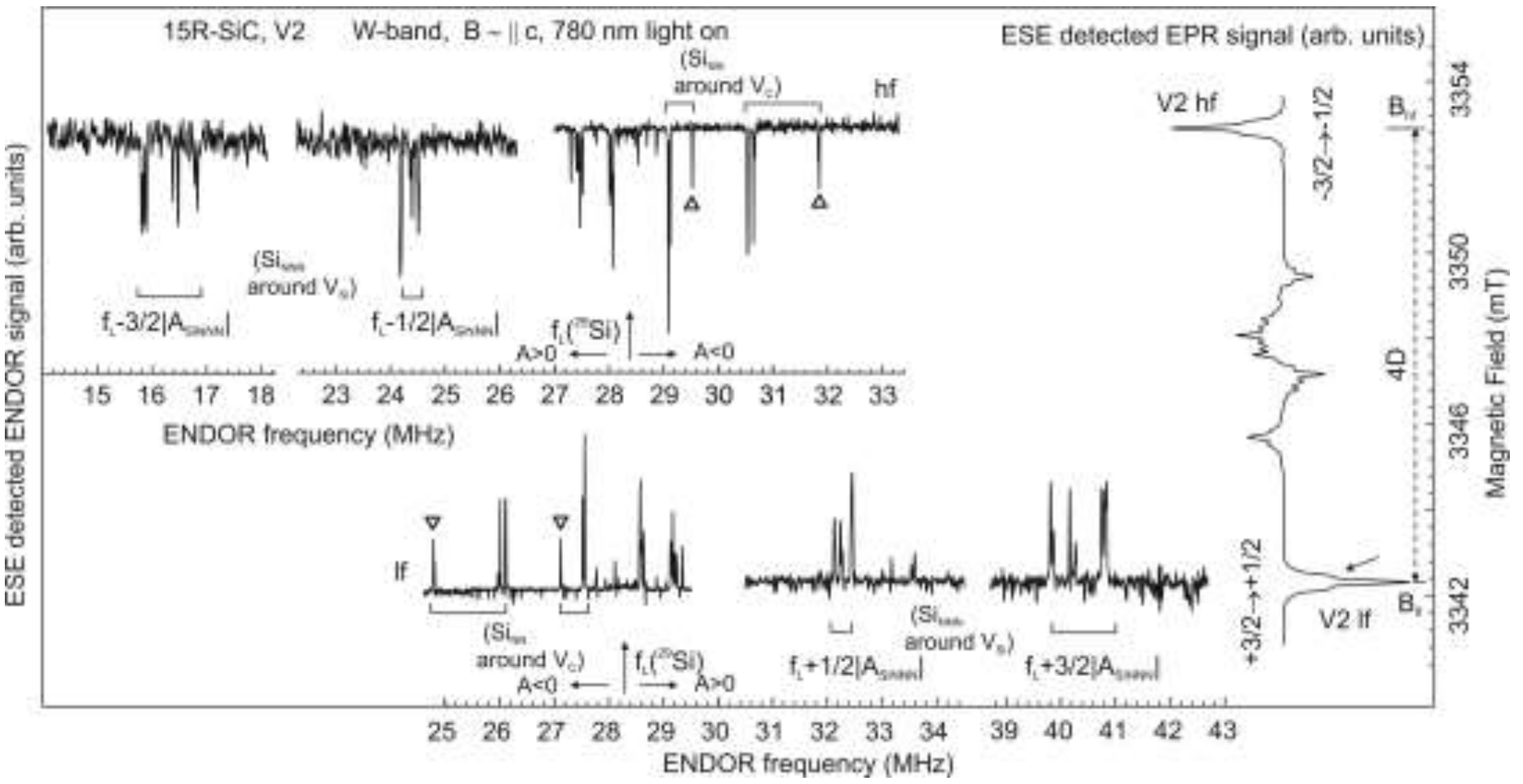
15R-SiC, V2
W-band, B ~ || c, 780 nm light on
ESE detected EPR signal (arb. units)
ESE detected ENDOR signal (arb. units)
ENDOR frequency (MHz)
Magnetic Field (mT)
V2 hf
V2 lf
hf
lf
4D
-3/2→-1/2
+3/2→+1/2
A>0
A<0

FIG. 7

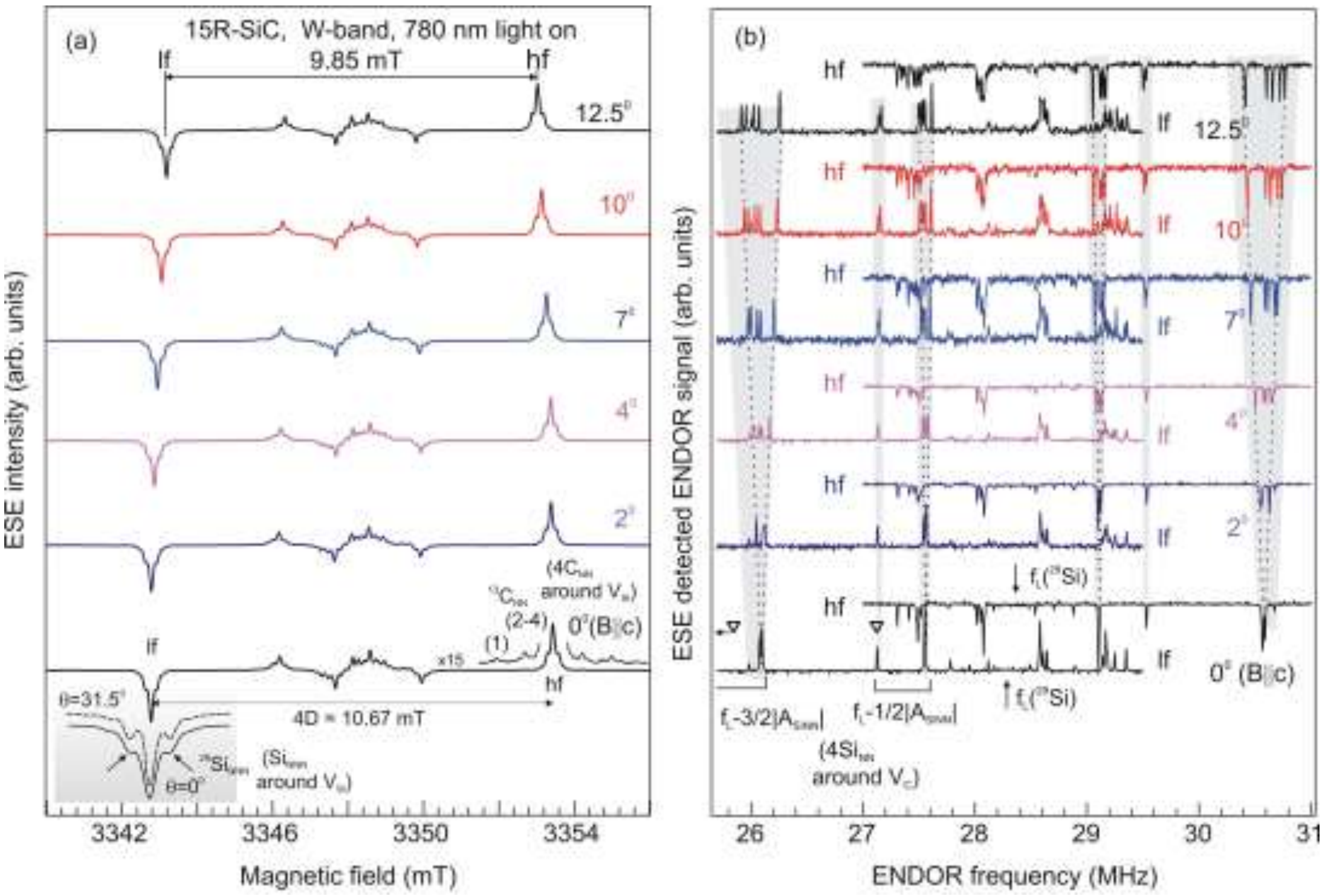
(a)
15R-SiC, W-band, 780 nm light on
9.85 mT
lf
hf
12.5°
10°
7°
4°
2°
0°(B||c)
4D = 10.67 mT
x15
ESE intensity (arb. units)
Magnetic field (mT)
(b)
ESE detected ENDOR signal (arb. units)
ENDOR frequency (MHz)

FIG. 8.

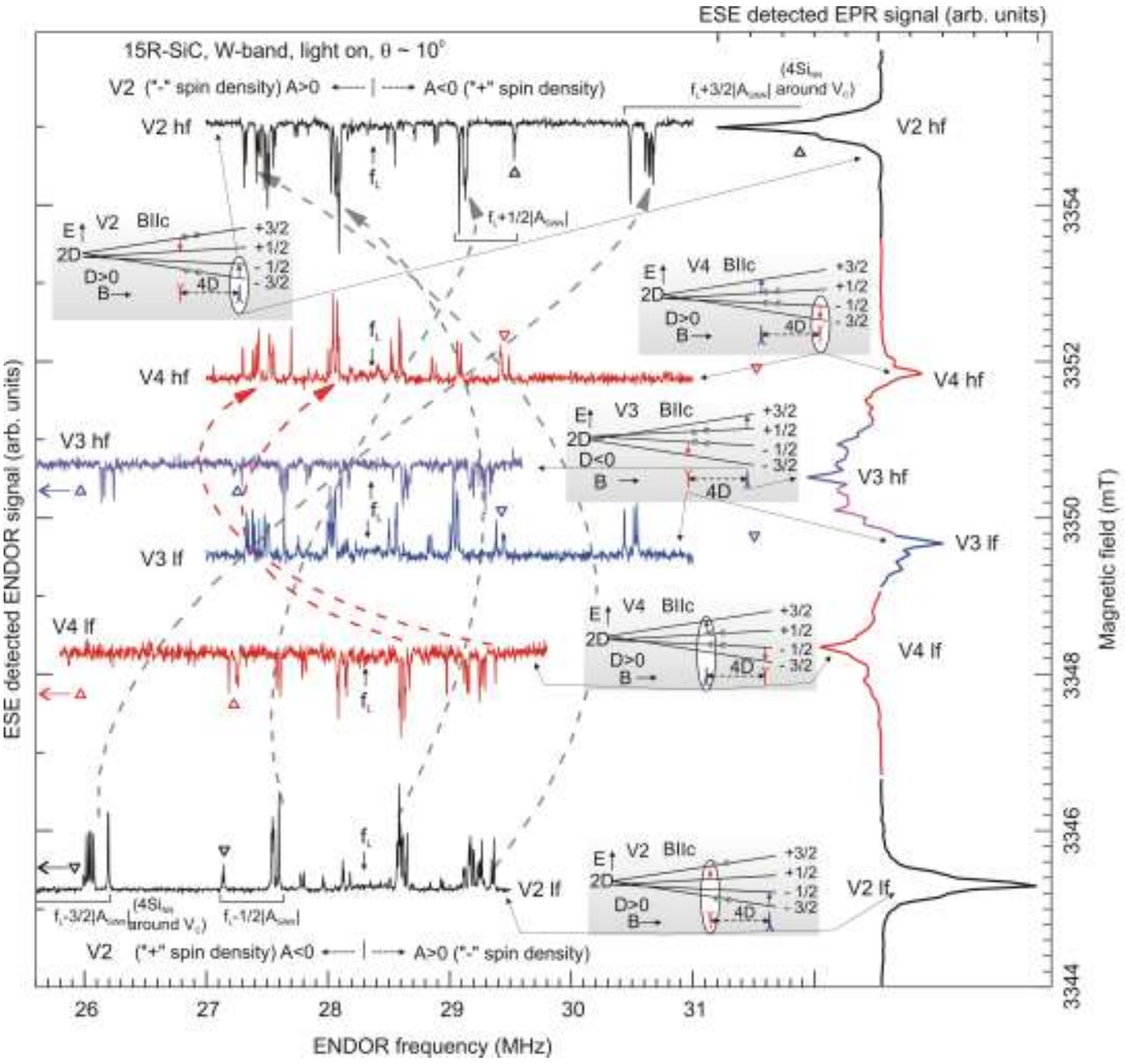

ESE detected EPR signal (arb. units)
15R-SiC, W-band, light on, θ ~ 10°
V2 ("-" spin density) A>0 | A<0 ("+" spin density)
V2 hf
V4 hf
V3 hf
V3 lf
V4 lf
V2 lf
V2 ("+" spin density) A<0 | A>0 ("-" spin density)
ESE detected ENDOR signal (arb. units)
ENDOR frequency (MHz)
Magnetic field (mT)
26
27
28
29
30
31
3344
3346
3348
3350
3352
3354

FIG. 9.

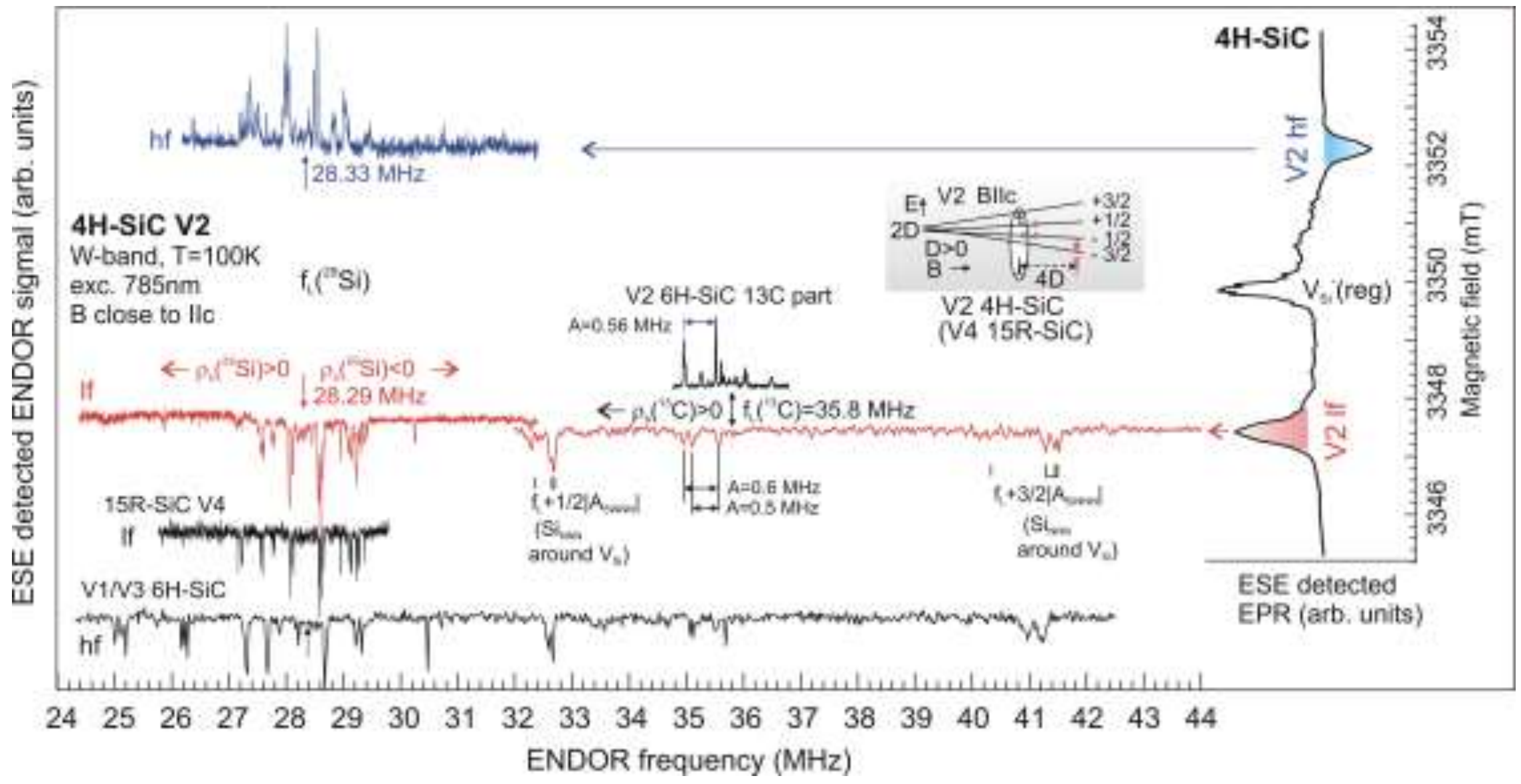

FIG. 10.

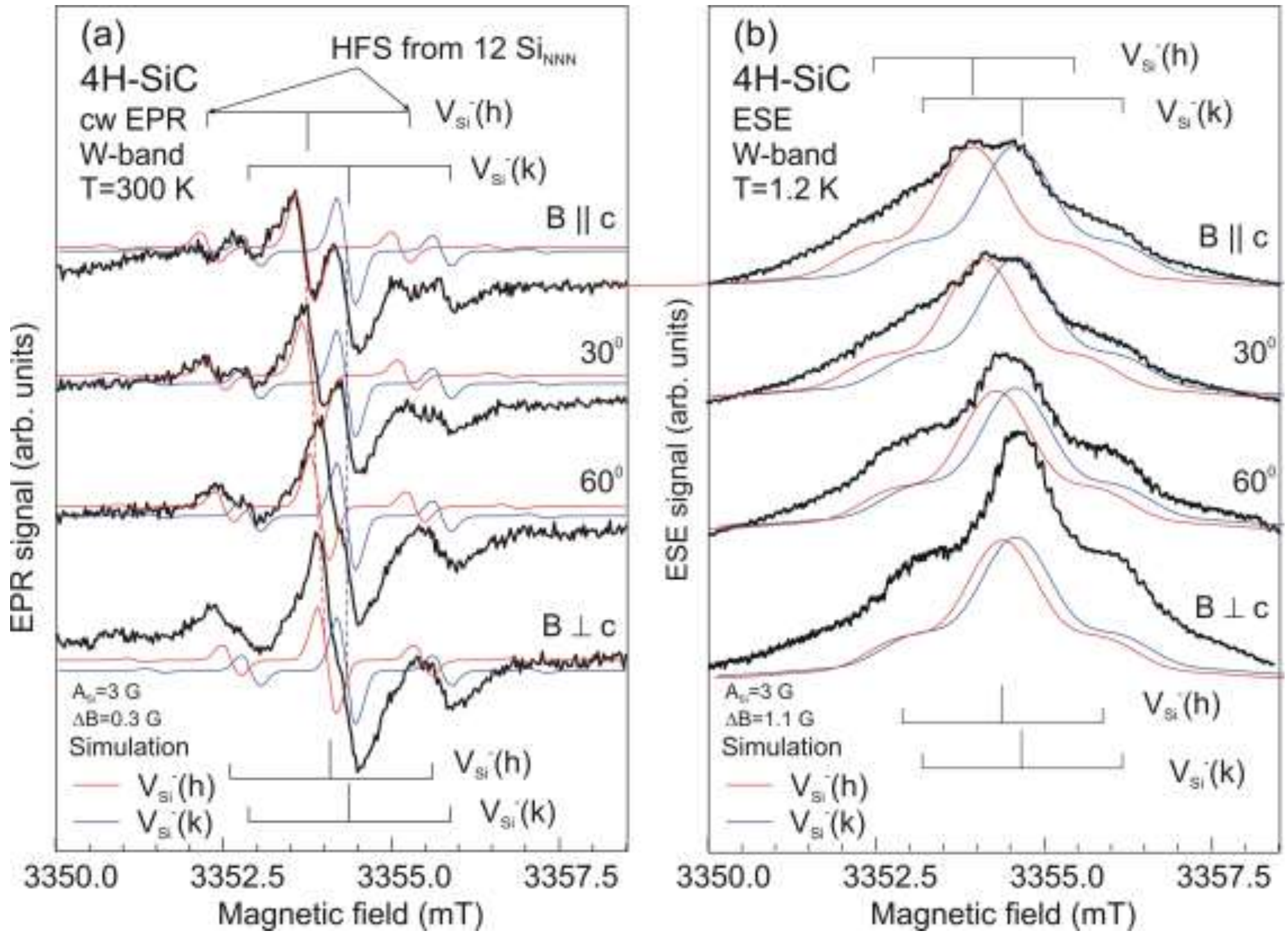

FIG. 11.

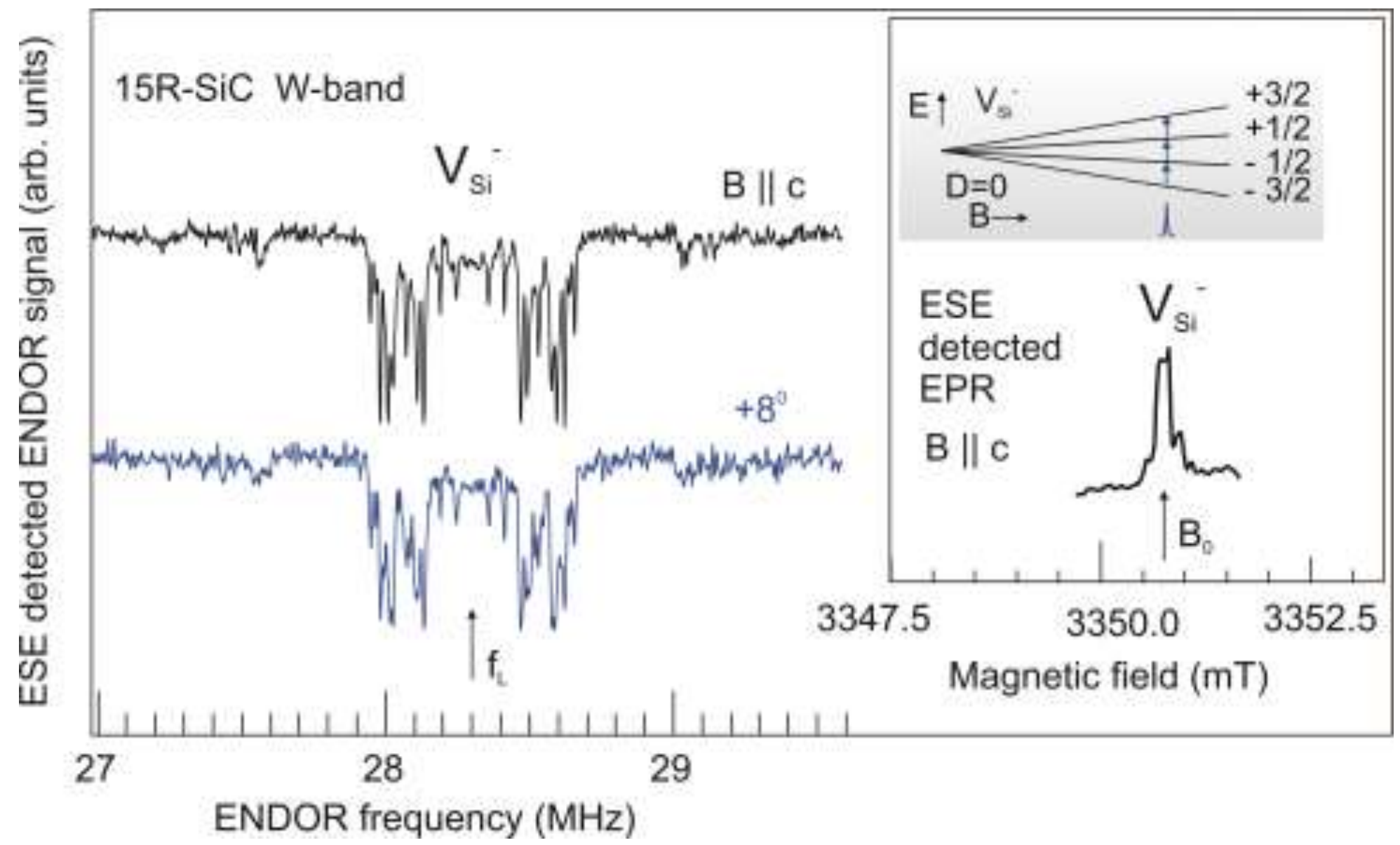

FIG. 12.

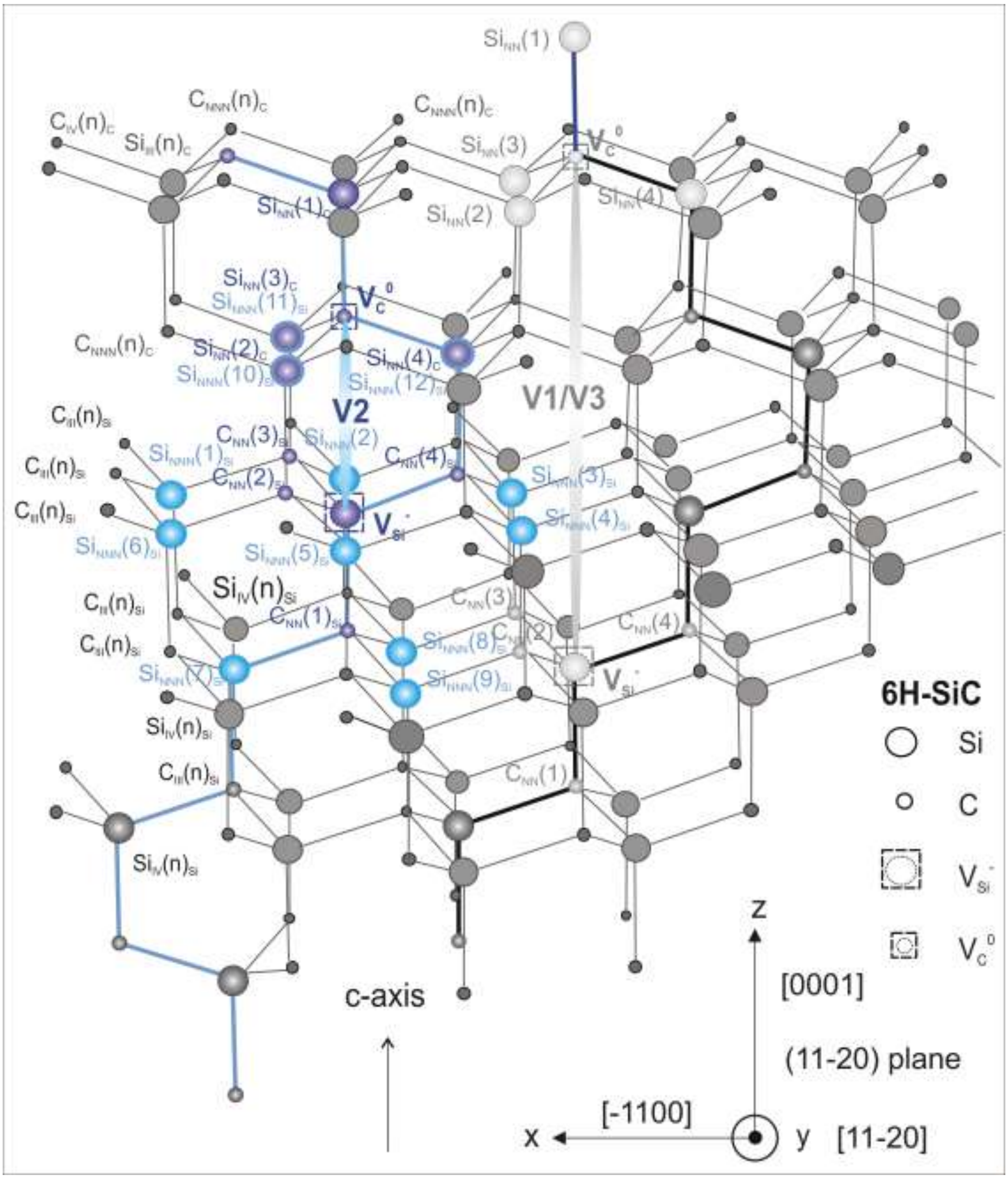

FIG. 13.

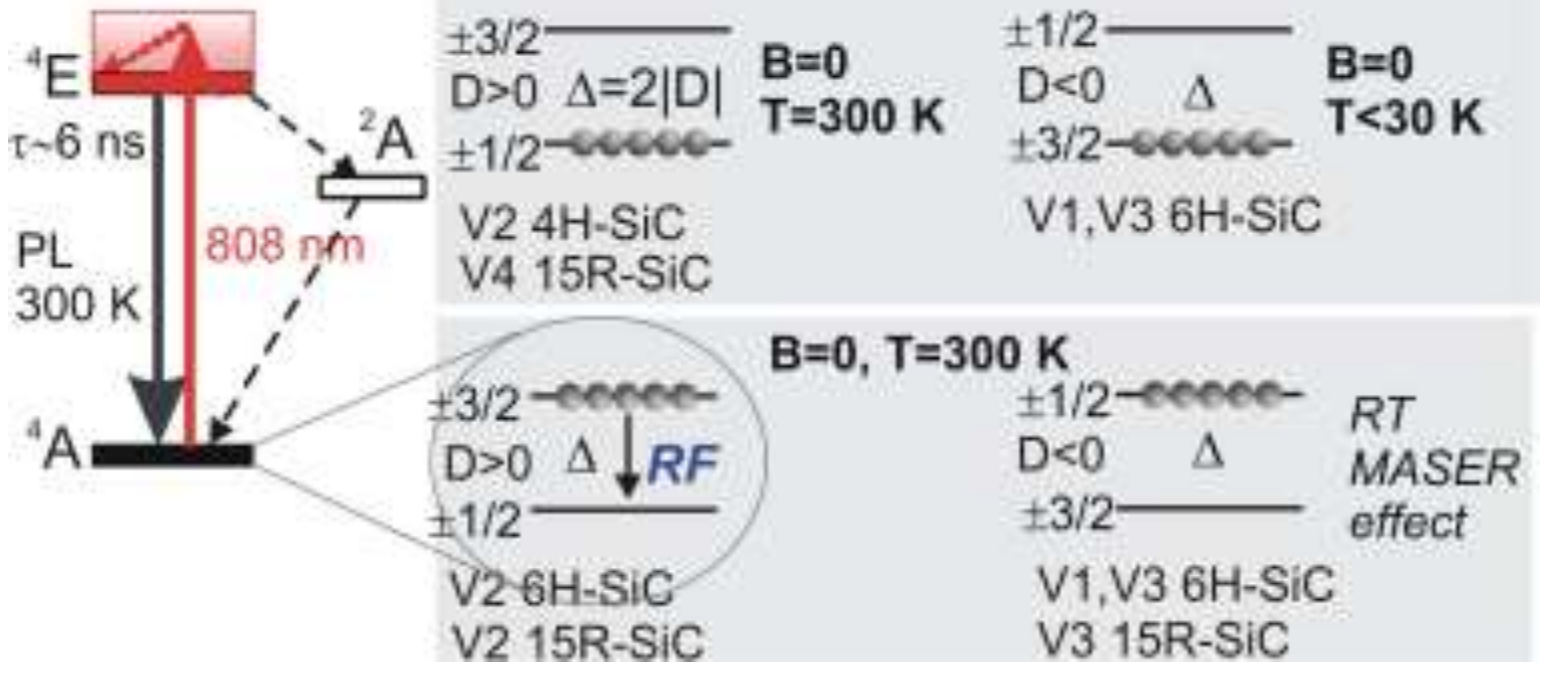

FIG. 14

TABLE II

Ligand hyperfine parameters for interaction with $^{13}C$-atoms and $^{29}Si$-atoms surrounding a spin color center of V1, V2, V3, V4 in three polytypes of SiC: 6H-, 15R- and 4H-SiC. The HF-tensor principal values *a* and *b* and the corresponding *s* and *p* spin densities of the unpaired electron connected to the spin centers in SiC with the nuclei surrounding the center. To consider the distribution of *s* and *p* character, the observed HF interactions were translated into spin density using the table of Morton and Preston [73]. The isotropic HF interaction (*a*) gives a measure of the s spin density (s), whereas the anisotropic HF interaction (*b*) is a measure for the *p* density (*p*). The HF interaction with four C-atoms located in the nearest-neighbor (NN) of a silicon vacancy (NN shell) has axial symmetry along the bonding direction, therefore parameters are given parallel to the bond ($A_{\|}$) and perpendicular to the bond ($A_{\perp}$), from EPR data [56]. The HF interaction with the twelve Si-atoms located in the next-nearest-neighbor (NNN) of a silicon vacancy (NNN shell) is almost isotropic (EPR data [56]).

| Crystal | Center | Atom | HF interaction for $^{29}Si$, $^{13}C$ (MHz) | *a* and *b* (MHz) | Spin density (%) |
|---|---|---|---|---|---|
| **6H-SiC** | V2 | $C_{NN}(V_{Si}^{-})_{1-4}$ From EPR [56] | along the c-axis (1) $A_{\|}$=80.1, $A_{\perp}$=37.5 off the c-axis (2-4) $A_{\|}$=80.1, $A_{\perp}$=30.8 | *a*=51.7 *b*=14.2 *a*=47.2 *b*=16.4 | $\rho_S$ >0 *s*=1.34 *p*=13.4 *s*=1.22 *p*=15.5 Σ=1.34+13.4+ (1.22+15.5)×3 =+64.9 |
| | | $Si_{NNN}$ ($V_{Si}^{-}$) | A =7.84 EPR [56] A≈*a*= 8.31, 7.92, 7.71 ENDOR | *a*=7.84 *a*=8.31, 7.92, 7.71≅8.0 | $\rho_S$ <0 s=-0.17 Σ=-0.17×12=-2.04 s= -0.18, -0.17, -0.165≅-0.17 Σ=-0.17×12=-2.04 |
| | | $C_{III}(V_{Si}^{-})$ | $A_{\|}$≅0.562, 0.542, $A_{\perp}$≅0.31 | *a*=0.39–0.38 *b*=0.087 | $\rho_S$ >0 s=0.01, p=0.08 |
| | | $Si_{IV}(V_{Si}^{-})$ | A≈*a*: -0.34, -0.22 | *a*≈ -0.34, -0.22 | $\rho_S$ >0 s=0.007, 0.005 |
| | | $Si_{NN}(V_C^0)_{1-4}$ | $A_{\parallel}$=-2.33; -2.32,<br>-2.31,<br>$A_{\perp}\cong$-1.42; -<br>1.40; -1.36 | *a*=-1.72, -<br>1.71, -1.68;<br>*b*=-0.3, -<br>0.31, -0.32 | $\rho_S$ >0<br>s=0.037÷<br>0.036;<br>p=0.26÷0.28 |
| | | $C_{NNN}(V_C^0)$ | -0.48<br>almost isotropic | | $\rho_S$ <0 |
| | | $Si_{III}(V_C^0)$ | $A_{\parallel}$=0.71<br>$A_{\perp}\cong$0.40<br>$A_{\parallel}$=0.65,<br>$A_{\perp}\cong$0.39<br>$A_{\parallel}$=0.61,<br>$A_{\perp}\cong$0.38<br>$A_{\parallel}$=0.58,<br>$A_{\perp}\cong$0.38<br>$A_{\parallel}$=0.57,<br>$A_{\perp}\cong$0.38 | a=0.5<br>b=0.1<br>a=0.48<br>b=0.09<br>a=0.46<br>b=0.08<br>a=0.45<br>b=0.07 | $\rho_S$ < 0<br>s=-0.011%÷<br>-0.010%,<br>p=-0.087%÷<br>-0.061% |
| | | $C_{IV}(V_C^0)$ | 0.278, 0.257 | | $\rho_S$ >0 |
| | | Remote Si shells | 0.26, 0.18, 0.12,<br>-0.05 | | $\rho_S$ <0<br>$\rho_S$ >0 |
| | | Remote C shells | -0.18<br>almost isotropic | | $\rho_S$ <0 |
| **6H-SiC** | V1/V3 | $C_{NN}\ (V_{Si}^-)_{1-4}$<br>From EPR [56] | V1<br>along the c-axis<br>(1) $A_{\parallel}$=71.7,<br>$A_{\perp}$=31.9<br>off the c-axis along the bonding direction (2-4)<br>$A_{\parallel}$=80.1,<br>$A_{\perp}$=30.2<br>V3<br>along the c-axis<br>(1) $A_{\parallel}$=80.1, | *a*=45.2<br>*b*=13.3<br>*a*=46.8<br>*b*=16.6<br><br>*a*=46.5 | $\rho_S$ >0<br>*s*=1.17 *p*=12.6<br>*s*=1.21 *p*=15.7<br>Σ=+64.5<br><br>*s*=1.20 *p*=15.9 |
| | | | $A_{\perp}$=29.7<br>off the c-axis along the bonding direction "z" (2-4)<br>$A_{\parallel z}$=74.8,<br>$A_{\perp y\langle 11\text{-}20\rangle}$=26.9,<br>$A_{\perp x}$=37.5, $A_{\perp}\cong$ $(A_{\perp y} + A_{\perp x})/2$ =32.2 | *b*=16.8<br>*a*=46.4<br>*b*=14.2 | *s*=1.20 *p*=13.4<br>Σ=+60.9 |
| | | $Si_{NNN}$ ($V_{Si}^{-}$) | V1<br>A =8.15<br>EPR [56]<br>V3<br>A =8.4<br>EPR [56]<br>A=8.58, 8.38<br>ENDOR | <br>*a*=8.15<br><br>*a*=8.4<br><br><br>*a* =8.58, 8.38 | $\rho_S$ <0<br>s=-0.21<br>Σ=-0.21x12=-2.52<br>s=-0.18<br>Σ=-0.18x12=-2.16<br>s=-0.22<br>s=-0.21<br>Σ=-[0.22x6+0.21x6]=-2.6 |
| | | $C_{III}$($V_{Si}^{-}$) | Not detected | | |
| | | $Si_{IV}$($V_{Si}^{-}$) | -0.47, -0.33<br>anisotropy less 0.5% | *a*≈ -0.47, -0.33 | $\rho_S$ >0<br>s=0.01, 0.007, |
| | | $Si_{NN}$($V_C^0$)$_{1-4}$ | $A_{\parallel}$=-2.22; -2.15<br>$A_{\perp}\cong$-1.35 | *a*=-1.64, -1.62;<br>b=-0.29, -0.27 | $\rho_S$ >0<br>s≅0.035;<br>p≅0.25 |
| | | $C_{NNN}$($V_C^0$) | -0.47<br>almost isotropic | *a*=-0.47 | $\rho_S$ < 0<br>s=-0.012 |
| | | $Si_{III}$($V_C^0$) | B‖c, $0^0$<br>0.664, 0.635, 0.567, 0.38 | | $\rho_S$ < 0 |
| | | | $20^0$<br>0.735, 0.702,<br>0.659, 0.629,<br>0.582, 0.548,<br>0.522, 0.493,<br>0.417, 0.371,<br>0.341, 0.313 | | |
| | | $C_{IV}(V_C^0)$ | -0.28 | | ρs <0 |
| | | Remote Si shells | -0.05 | | ρs >0 |
| | $V_{Si}^-$ | $C_{NN}$ $(V_{Si}^-)_{1-4}$<br>From EPR [56] | along the c-axis<br>(1) $A_{\parallel}$=80.4,<br>$A_{\perp}$=32.2<br>off the c-axis<br>(2-4) $A_{\parallel}$=80.4,<br>$A_{\perp}$=32.2 | a=48.3<br>b=16.1<br>a=48.3<br>b=16.1 | ρs >0<br>*s*=1.24 *p*=15.2<br>*s*=1.24 *p*=15.2<br>Σ=+65.8 |
| | | $Si_{NNN}$ $(V_{Si}^-)$<br>From EPR [56] | A =8.32 | *a*=8.32 | ρs <0<br>*s*=-0.214<br>Σ=-2.57 |
| | | | | | |
| **15R-SiC** | V2 | $C_{NN}(V_{Si}^-)_{1-4}$ | $A_{\parallel}$=84.6<br>$A_{\perp}$=33.6 | *a*=50.6<br>*b*=17.0 | ρs >0<br>*s*=1.31<br>*p*=16.1<br>Σ≈ (1.31+16.1)×4<br>=+69.6 |
| | | $Si_{NNN}$ $(V_{Si}^-)$ | A≈*a*: 8.35, 8.03, 7.92, 7.75, 7.69<br>almost isotropic | *a* ≈ 8.35, 8.03, 7.92, 7.75, 7.69 | ρs <0<br>s= -0.18, -0.17, -0.165≅-0.17<br>Σ=-0.17×12=-2.04 |
| | | $C_{III}(V_{Si}^-)$ | Not detected | | ρs >0 |
| | | $Si_{IV}(V_{Si}^-)$ | A≈*a*: -0.34, -0.23<br>almost isotropic | *a* ≈-0.34, -0.23 | ρs >0<br>s=0.007, 0.005 |
| | | $Si_{NN}(V_C^0)_{1-4}$ | $A_{\parallel}\cong$-2.34; $A_{\perp}\cong$-1.43, 1.40 | *a*=-1.72, -1.71; *b*=-0.3, -0.31 | $\rho_S$ >0 s=0.036; p=0.26÷0.27 |
| | | $C_{NNN}(V_C^0)$ | Not detected | | |
| | | $Si_{III}(V_C^0)$ | $A_{\parallel}$=0.71, $A_{\perp}\cong$0.42 $A_{\parallel}$=0.64, $A_{\perp}\cong$0.42 $A_{\parallel}$=0.58, $A_{\perp}\cong$0.41 $A_{\parallel}$=0.57, $A_{\perp}\cong$0.41 | a=0.52 b=0.097 a=0.49 b=0.07 a=0.47 b=0.06 a=0.47 b=0.06 | $\rho_S$ < 0 s=-0.011÷ -0.010, p=-0.087÷ -0.061 |
| | | $C_{IV}(V_C^0)$ | Not detected | | |
| | | Remote Si shells | 0.26, 0.18, 0.12 | | $\rho_S$ < 0 |
| | V3 | $C_{NN}$ $(V_{Si}^-)_{1-4}$ | From EPR approximately as for V2 | | |
| | | $Si_{NNN}$ $(V_{Si}^-)$ | From EPR approximately as for V2 | | |
| | | $C_{III}(V_{Si}^-)$ | Not detected | | |
| | | $Si_{IV}(V_{Si}^-)$ | A≈*a*: -0.44, -0.33, -0.32 | *a*≈ -0.44, -0.33, -0.32 | $\rho_S$ >0 s=0.009, 0.007 |
| | | $Si_{NN}(V_C^0)_{1-4}$ | ~$10^0$ -2.33, -2.21, -2.13, -1.49, -1.46, -1.41 ($A_{\parallel}\cong$-2.22, - | *a*≅-1.67, -1.6; -0.27, -0.25 | $\rho_S$ >0 s≅0.035 p≅0.23 |
| | | | 2.20, -2.10; $A_{\perp}$≅-1.40, -1.35) | | |
| | | $C_{NNN}(V_C^0)$ | Not detected | | |
| | | $Si_{III}(V_C^0)$ | ~$10^0$ 0.68, 0.66, 0.64, 0.62, 0.59, 0.55, 0.54, 0.51, 0.4, | | $\rho_S < 0$ |
| | | $C_{IV}(V_C^0)$ | Not detected | | |
| | | Remote Si shells | 0.2 | | $\rho_S < 0$ |
| | V4 | $C_{NN}$ $(V_{Si}^-)_{1\text{-}4}$ | From EPR approximately as for V2 | | |
| | | $Si_{NNN}$ $(V_{Si}^-)$ | From EPR approximately as for V2 | | |
| | | $C_{III}(V_{Si}^-)$ | Not detected | | |
| | | $Si_{IV}(V_{Si}^-)$ | ~$10^0$ -0.50, -0.47, -0.34, -0.32, | | $\rho_S > 0$ |
| | | $Si_{NN}(V_C^0)_{1\text{-}4}$ | ~$10^0$ -2.3, -2,13, -2.11 | | $\rho_S >0$ |
| | | $C_{NNN}(V_C^0)$ | Not detected | | |
| | | $Si_{III}(V_C^0)$ | ~$10^0$ 1.3, 0.70, 0.67, 0.65, 0.62, 0.55, 0.53, | | $\rho_S <0$ |
| | | $C_{IV}(V_C^0)$ | Not detected | | |
| | | Remote Si shells | 0.28, 0.12 | | $\rho_S <0$ |
| | $V_{Si}^-$ | $C_{NN}$ $(V_{Si}^-)_{1-4}$ | From EPR approximately as for V2 | | |
| | | $Si_{NNN}$ $(V_{Si}^-)$ | From EPR approximately as for V2 | | |
| | | $C_{III}(V_{Si}^-)$ | Not detected | | |
| | | $Si_{IV}(V_{Si}^-)$ | 0.24÷0.12<br>0.58÷0.48 | | |
| **4H-SiC** | V1 | $C_{NN}$ $(V_{Si}^-)_{1-4}$<br>From EPR [56] | along the c-axis<br>(1) $A_{\|\|}$=71.7, $A_{\perp}$=31.9<br>off the c-axis<br>(2-4) $A_{\|\|}$=78.4, $A_{\perp}$=31.4 | *a*=45.2<br>*b*=13.3<br>*a*=47.1<br>*b*=15.7 | $\rho_S$ >0<br>*s*=1.17<br>*p*=12.6<br>*s*=1.22<br>*p*=14.9<br>Σ=+62.1 |
| | | $Si_{NNN}$ $(V_{Si}^-)$<br>From EPR [56] | A =8.15 | *a*=8.15 | $\rho_S$ <0<br>*s=-0.21*<br>Σ=0.21x12=-2.52 |
| | V2 | $C_{NN}$ $(V_{Si}^-)_{1-4}$<br>From EPR [56] | along the c-axis<br>(1) $A_{\|\|}$=82.9, $A_{\perp}$=34.7<br>off the c-axis<br>(2-4) $A_{\|\|}$=75.6, $A_{\perp}$=29.1 | *a*=50.8<br>*b*=16.1<br>*a*=44.6<br>*b*=15.5 | $\rho_S$ >0<br>*s*=1.31 *p*=15.2<br>*s*=1.15 *p*=14.6<br>Σ=+63.8 |
| | | $Si_{NNN}$ $(V_{Si}^-)$ | A =8.34<br>EPR [56]<br><br>A = 8.81, 8.74, 8.62, 7.90<br>ENDOR | *a*=8.34<br><br><br>*a* =8.81, 8.74, 8.62, 7.90 | $\rho_S$ <0<br>*s*=-0.215<br>Σ=0.215x12=-2.58.<br>*s*=-0.23, -0.225, -0.22,<br>-0.20 |
| | | $C_{III}(V_{Si}^-)$ | Not detected | | |
| | | $Si_{IV}(V_{Si}^-)$ | A=-0.5, -0.468, -0.447, -0.367, -0.32 | *a*≈-0.5, -0.468, -0.447, - | $\rho_S$ >0<br>s=0.01÷ 0.007 |
| | | | | 0.367, -0.32 | |
| | | $Si_{NN}(V_C^0)_{1-4}$ | $A_{\parallel}$=-2.3, -2.14<br>$A_{\perp}$=-1.5 | a=-1.77, -1.71<br>b=-0.27, -0.21 | $\rho_S$ >0<br>s=0.037÷0.036;<br>p=0.23 |
| | | $C_{NNN}(V_C^0)$ | Not detected | | |
| | | $Si_{III}(V_C^0)$ | A=1.307, 1.292, 0.743, 0.719, 0.702, 0.671, 0.620, 0.605, 0.569, 0.524, 0.521, | | $\rho_S$ <0 |
| | | $C_{IV}(V_C^0)$ | Not detected | | |
| | | Remote Si shells | 0.11 | | $\rho_S$ <0 |
| | $V_{Si}^-$<br>g(k)=2.0032<br>$g_{\parallel}$(h)=g(k)+0.00004<br>$g_{\perp}$(h)=g(k)+0.00002 | $C_{NN}$ $(V_{Si}^-)_{1-4}$<br>From EPR [56] | along the c-axis (1) $A_{\parallel}$=80.1, $A_{\perp}$=33.9<br>off the c-axis (2-(4) $A_{\parallel}$=80.1, $A_{\perp}$=33.9 | *a*=49.3<br>*b*=15.4<br>*a*=49.3<br>*b*=15.4 | $\rho_S$ >0<br>*s*=1.27 *p*=14.5<br>*s*=1.27 *p*=14.5<br>Σ=+63.1 |
| | | $Si_{NNN}$ $(V_{Si}^-)$<br>From EPR [56] | A =8.34 | *a*=8.34 | $\rho_S$ <0<br>*s*=-0.215<br>Σ=-0.215x12=-2.58 |